# ArgRE: Formal Argumentation for Conflict Resolution in Multi-Agent Requirements Negotiation

HAOWEI CHENG[1], MILHAN KIM[1], CHONG LIU [2], TRUONG VINH TRUONG DUY[4], PHAN THI HUYEN THANH[4], TEERADAJ RACHARAK[3], JIALONG LI[1], NAOYASU UBAYASHI[1], and HIRONORI WASHIZAKI[1],(Senior Member, IEEE)
[1]Department of Computer Science and Communications Engineering, Waseda University, Tokyo, Japan
[2]College of Architecture and Urban Planning, Tongji University, Shanghai, China
[3]Advanced Institute of So-Go-Chi (Convergence Knowledge) Informatics, Tohoku University, Miyagi, Japan
[4]Aisin Corporation, Tokyo, Japan

Corresponding author: Haowei Cheng (e-mail: haowei.cheng@fuji.waseda.jp).

**ABSTRACT** As software systems grow in complexity, they must satisfy an increasing number of competing quality attributes, making it essential to balance them in a principled manner—for example, a safety requirement for sensor-fusion verification may conflict with a tight planning-cycle budget. Multi-agent large language model frameworks support this balancing process by assigning specialized agents to different objectives. However, their conflict resolution is typically heuristic. Requirements are aggregated implicitly without explicit acceptance or rejection, limiting auditability in regulated domains. We present ArgRE, a multi-agent requirements negotiation system that embeds Dung-style abstract argumentation into the negotiation stage. Each proposal, critique, and refinement is modeled as an argument, conflicts are represented as directed attack relations, and the accepted set of arguments is computed under grounded and preferred semantics. The pipeline further integrates KAOS goal modeling, multi-layer verification, and standards-oriented artifact generation. Evaluation across five case studies spanning safety-critical, financial, and information-system domains shows that ArgRE provides argument-level traceability absent from existing frameworks. Independent evaluators rated its decision justifications significantly higher than those of heuristic synthesis (4.32 vs. 3.07, $p < 0.001$), indicating improved auditability, while semantic intent preservation remains comparable (94.9% BERTScore F1) and compliance coverage reaches 84.7% versus 47.6%–47.8% for baselines. Structural analysis further confirms that the default pairwise protocol yields acyclic graphs in which grounded and preferred semantics coincide, whereas cross-pair arbitration introduces controlled cyclicity, leading to predictable divergence between the two semantics.

## I. INTRODUCTION

Requirements Engineering (RE) is the earliest phase of the software development lifecycle and critically influences the quality and success of the final product [1]. For example, the perception subsystem of an autonomous vehicle must simultaneously satisfy stringent safety verification requirements and real-time latency constraints—objectives that often conflict and require transparent justification under standards such as ISO 26262. With the proliferation of autonomous systems and AI-driven applications, RE has evolved from focusing on functional specifications to addressing a multi-dimensional spectrum of quality attributes [2]. Modern software systems must satisfy competing non-functional requirements, including safety, efficiency, environmental sustainability, and trustworthiness [3]. As system complexity increases, manually balancing these dimensions while maintaining consistency becomes error-prone and labor-intensive [4].

Despite decades of research, RE remains a major source of software project failure. Empirical studies indicate that over 70% of failed projects can be attributed to requirements-related issues, including incompleteness, ambiguity, and unresolved stakeholder conflicts [5], [6]. These challenges are further intensified in safety-critical domains such as autonomous vehicles and medical devices, where systems must simultaneously satisfy stringent and often competing quality attributes [7].

Recent advances in large language models (LLMs) have spurred increasing interest in automating RE tasks [8], [9]. Multi-agent frameworks such as MARE [10], iReDev [11], and other quality-oriented systems demonstrate that end-to-end automation can be achieved through role decomposition and collaborative orchestration.

However, a fundamental limitation persists: conflict resolution remains predominantly heuristic. Requirements are typically merged through implicit aggregation, coordinator-mediated moderation, or priority-weighted synthesis, without formal acceptability semantics (i.e., Dung-style argumentation semantics [12]). Although often effective in practice, these mechanisms do not provide formal justification for why specific requirements are retained, revised, or rejected under multi-party quality conflicts. This lack of justification limits interpretability, auditability, and theoretical grounding—properties essential in regulated engineering contexts such as the EU AI Act (Article 11) [13] and ISO 26262 [14] workflows.

Why argumentation rather than multi-objective optimization? Optimization techniques (e.g., Pareto-front analysis) identify optimal trade-off solutions but do not inherently capture the rationale behind selecting one trade-off over another. In regulated domains, the primary requirement is not merely to identify an optimal balance, but to provide an auditable trace explaining, for each accepted requirement, which competing concerns were raised, how they were addressed, and under which acceptance criterion the final formulation was retained. Argumentation frameworks (AFs) are explicitly designed for this purpose: they represent the structure of conflicts among competing claims and compute accepted sets under transparent, inspectable semantics, where an argument is considered *defended* if each of its attackers is counterattacked by an accepted argument. The effectiveness of AFs for conflict resolution is well established and continues to be an active area of research. Since Dung's seminal proposal, argumentation has been applied to design rationale [15], architectural decision-making [16], and requirements rationale management [17].

More recently, [18] showed that fine-tuning LLMs on abstract argumentation benchmarks significantly improves their conflict-resolution capabilities, indicating that AF-based semantic computation transfers effectively to LLM-driven settings. This finding motivates ArgRE's integration of Dung-style semantics into an LLM-based multi-agent pipeline. AFs are therefore well suited to multi-quality negotiation, where transparency and traceability are primary objectives, including trade-offs involving efficiency, sustainability, and other non-safety quality dimensions increasingly subject to regulatory scrutiny (e.g., energy-efficiency reporting under ISO 14001 and the EU AI Act's transparency obligations).

We term this limitation the *Formalization Gap*: the absence of a mathematically grounded decision framework that can justify the acceptance or rejection of requirements under multi-party conflicts with competing quality objectives. Addressing this gap requires well-defined acceptability criteria and deterministic resolution rules, rather than implicit aggregation or ad hoc priority weighting. As noted above, *formal* in this paper refers to Dung's abstract argumentation semantics, in which acceptability is determined by attack relations and the defense properties they induce, under grounded and preferred-extension criteria.

To illustrate the unresolved conflicts arising from this gap, consider an autonomous-driving perception module in which a Safety agent proposes a 500-ms sensor-fusion verification budget, while an Efficiency agent requires fusion within 30 ms. Under heuristic synthesis, such constraints are typically combined into a compromise value without recording which concern was prioritized. In contrast, an argumentation-based approach models each critique and refinement as an explicit attack or defense, yielding a traceable justification for every accepted requirement. Section IV-H traces this conflict through the full ArgRE pipeline.

To close this gap, we propose ArgRE, a multi-agent RE system centered on argumentation-based conflict resolution. ArgRE coordinates quality-specialized agents grounded in ISO/IEC 25010, conducts dialectical negotiation to surface conflicts, and integrates a Dung-style argumentation layer with KAOS-based specification, verification, and compliance-oriented artifact generation. This design leverages a natural correspondence between multi-quality conflicts and argumentation structure: when agents advocating different quality objectives (e.g., safety versus efficiency, or performance versus sustainability) produce competing requirements, each critique or counter-proposal maps to an *attack* in Dung's framework, while each refinement corresponds to a counter-attack that reinstates the defended proposal under Dung's semantics. The resulting attack graph provides a precise, inspectable record of how cross-quality trade-offs are negotiated and resolved, making acceptability explicit end-to-end rather than relying on opaque, coordinator-driven synthesis. Accordingly, instead of relying on implicit synthesis alone, ArgRE:

1) Models requirement proposals, critiques, and refinements as *arguments*.
2) Represents conflicts among requirements as *attack relations*.
3) Computes accepted requirement sets under formal *argumentation semantics* (grounded and preferred).
4) Provides full *provenance traces* from accepted requirements back to originating arguments.

The resulting pipeline makes conflict resolution explicit and traceable, linking accepted requirements to their argument-level origins under the chosen argumentation semantics.

The contributions of this paper are as follows:

- **An end-to-end framework with argumentation-grounded provenance for conflict resolution.** We present ArgRE, a multi-agent requirements negotiation system that replaces heuristic synthesis with Dung-style argumentation semantics. Each accepted requirement is accompanied by a complete provenance trace, from its originating proposal through intermediate critiques and refinements, with attack relations defined by three deterministic structural patterns and complemented by a threshold-gated LLM pathway. This provenance capability is not provided by existing multi-agent RE frameworks and directly supports auditability mandates in regulated contexts (EU AI Act Article 11, ISO 26262). The pipeline further integrates KAOS goal modeling, multi-layer verification, and standards-oriented artifact generation.
- **Empirical demonstration that formal argumentation preserves semantic intent while enhancing interpretability.** Across five case studies and 30 runs, ArgRE achieves semantic preservation (94.9% BERTScore) statistically indistinguishable from its heuristic ablation, while providing 40.9% average Trace Completeness (TC) and significantly higher Decision Justification Scores (4.32 vs. 3.07, $p < 0.001$). These results indicate that the argumentation layer functions as a transparent overlay, adding attack–defense structure and traceability without distorting negotiated intent.
- **Structural characterization of argumentation graphs with auditability-oriented metrics.** We introduce TC, Decision Justification Score (DJS), and Graph Cyclicity Index (GCI) to evaluate transparency and structural properties of the argumentation layer. GCI serves as a diagnostic indicator of semantics divergence: under the pairwise protocol ($GCI = 0$), grounded and preferred semantics coincide; under cross-pair arbitration ($GCI \approx 0.25$), they diverge with measurable downstream effects. A threshold sensitivity analysis shows that this transition follows a predictable cascade, from additional attack edges to increased cyclicity and subsequent semantics divergence.

*Pipeline overview.*
ArgRE follows a five-phase pipeline: Phase 1 (*Parallel Generation*) elicits requirements from multiple quality perspectives; Phase 2 (*Dialectical Negotiation*) resolves conflicts under argumentation semantics; Phase 3 (*KAOS Integration*) organizes accepted requirements into a goal hierarchy; Phase 4 (*Verification*) checks structural and compliance properties; and Phase 5 (*Output Generation*) produces auditable deliverables. The primary contribution lies in Phase 2, which replaces heuristic synthesis with Dung-style argumentation. Section IV describes each phase in detail.

The remainder of this paper is organized as follows. Section II reviews related work. Section III presents the theoretical background on formal argumentation. Section IV describes the ArgRE system design. Section V outlines the experimental setup. Section VI reports the results. Section VII discusses the implications. Section VIII addresses threats to validity (including methodological limitations), and Section IX concludes the paper.

## II. RELATED WORK
### A. LLM-BASED REQUIREMENTS ENGINEERING
Recent advances in LLMs have renewed interest in automating RE tasks, including requirements elicitation, classification, refinement, and specification generation [8], [9]. Elicitron [19] employs a population of LLM-generated user agents to simulate diverse perspectives and uncover latent needs; however, these agents operate independently without inter-agent negotiation. Most LLM-based RE approaches rely on single-agent reasoning [20], [21], in which a single model performs all analysis within a unified prompt, implicitly optimizing for linguistic plausibility rather than explicit multi-objective trade-offs. Cheng et al. [22] provide a comprehensive survey of generative AI in RE. Earlier foundational work, including the roadmap by Nuseibeh and Easterbrook [23] and elicitation surveys by Zowghi and Coulin [24], established the methodological basis upon which these LLM-based approaches build.

### B. MULTI-AGENT SYSTEMS IN REQUIREMENTS ENGINEERING
Collaborative RE has long pursued structured approaches to negotiation. Early work such as Xipho [25] models requirements analysis as interactions among autonomous agents, while the WinWin-based negotiation methodology of Grünbacher and Seyff [26] shows that stakeholder negotiation can surface conflicts and improve consistency through systematic resolution mechanisms. MARE decomposes RE into elicitation, modeling, verification, and specification tasks, with task-specialized agents collaborating through a shared workspace. iReDev extends this approach with knowledge-driven agents and a human-in-the-loop mechanism for artifact refinement. Both frameworks demonstrate the feasibility of end-to-end automation but primarily emphasize task orchestration rather than negotiation across quality dimensions. Neither framework explicitly models negotiation dynamics among competing quality attributes, leaving the rationale underlying trade-off decisions opaque. Cheng et al. [27] propose QUARE, which grounds agent roles in ISO/IEC 25010 and introduces dialectical negotiation. However, its conflict resolution remains heuristic, relying on priority-weighted synthesis without formal acceptability guarantees.

### C. ARGUMENTATION THEORY IN SOFTWARE ENGINEERING
Formal argumentation has a long history in AI and knowledge representation. The seminal work of Dung [12] introduced abstract AFs, establishing attack-based semantics for argument acceptability. Subsequent extensions include value-based AFs (VAFs) [28], which associate arguments with values and resolve conflicts based on value preferences, and bipolar AFs (BAFs) [29], which incorporate both attack and support relations.

In software engineering, argumentation has been applied to design rationale [15], architectural decision-making [16], and requirements rationale management [17]. Murukannaiah et al. [25] incorporate argumentation concepts into agent-oriented RE, while Mirbel and Villata [30] explore its use for requirements validation. More recent work by Racharak bridges formal argumentation and natural-language argument extraction. Racharak [31] applies Dung-style abstract argumentation to extract and evaluate arguments from natural-language product reviews, computing admissible sets from user-generated text. Racharak et al. [32] combine assumption-based argumentation with LLM-based prompt engineering to extract argumentative structures from hotel reviews. However, these applications focus on opinion analysis in product and hotel reviews rather than multi-agent RE negotiation, and do not address extraction from structured negotiation logs or scalable attack-graph construction across competing quality dimensions in an RE context. Walton and Krabbe [33] characterize dialogue types, including persuasion and negotiation, providing a foundation for structured requirements dialogue. More recent work on LLM-based multi-agent debate [34]–[36] shows that structured disagreement improves output quality compared to single-agent self-critique. However, these approaches lack formal acceptability semantics, leading to two key challenges for RE practitioners in regulated domains. First, without a formally defined accepted set, there is no principled answer to the question "why was this requirement retained and that one rejected?"—a question routinely raised by auditors under ISO 26262 and EU AI Act Article 11. Second, prior applications of argumentation in software engineering operate on isolated artifacts (e.g., a single product review or design decision) rather than on multi-round, multi-party negotiation logs, where competing quality dimensions produce interleaved proposals, critiques, and refinements. Applying these techniques directly would require: (i) extracting structured arguments from free-form, multi-agent negotiation transcripts; (ii) constructing attack relations that span quality dimensions and negotiation rounds; and (iii) scaling extension computation to graphs with tens of arguments and cross-pair conflicts. These challenges are not addressed by existing opinion-mining or design-rationale approaches.

ArgRE addresses both challenges by integrating Dung-style argumentation into an LLM-based multi-agent RE pipeline. Arguments are derived from structured negotiation logs using three deterministic patterns: critique attacks proposal, refinement attacks original, and refinement attacks critique, denoted as critique→proposal, refinement→original, and refinement→critique, where → indicates the direction of attack. A threshold-gated LLM pathway further captures cross-pair semantic conflicts. The resulting attack graph enables formal extension computation, and each accepted requirement retains a provenance trace to its originating quality objective. Table 1 summarizes the positioning of ArgRE relative to existing approaches across key dimensions.

## III. BACKGROUND

### A. KAOS MODELING CONTEXT

ArgRE requires a goal-modeling language that satisfies three criteria:

(i) a hierarchical refinement structure capable of representing requirements across multiple abstraction levels (Strategic, Tactical, and Operational);

(ii) formal refinement links enabling automated consistency checking; and

(iii) established tool support for downstream verification and compliance artifact generation. KAOS [37] satisfies these requirements. Its directed goal hierarchy, with explicit AND/OR refinement links, provides a structurally sound foundation for organizing negotiated requirements. In addition, its formal specification capabilities—specifically temporal-logic annotations on goals—support the deterministic verification performed in Phase 4[1]. Alternative goal-modeling frameworks were also evaluated. i* [38] emphasizes actor-dependency modeling, which is valuable for stakeholder analysis, however, it lacks the hierarchical refinement structure required for multi-level requirement decomposition and compliance mapping. GRL (Goal-oriented Requirement Language) [39] supports contribution links and satisfaction levels but does not provide the formal specification mechanisms necessary for rule-based consistency checking in Phase 4. SysML requirement diagrams offer traceability but do not natively support goal refinement hierarchies.

Taken together, these comparisons justify the selection of KAOS as the goal-modeling backbone for ArgRE. We next clarify its role within the overall pipeline. In ArgRE, KAOS integration serves two primary functions: (1) transforming the accepted requirement set $\mathcal{R}^{\text{acc}}$ from Phase 2 into a structurally valid goal graph with explicit parent–child refinement links; and (2) supporting downstream verification (Phase 4) and standards-oriented artifact generation (Phase 5) based on a consistent, validated goal topology.

### B. ARGUMENTATION FRAMEWORKS

#### 1) Framework of Dung

An abstract AF [12] is defined as a pair:

$$AF = \langle \mathcal{A}, \mathcal{R}_{att} \rangle, \tag{1}$$

where $\mathcal{A}$ is a finite set of arguments and $\mathcal{R}_{att} \subseteq \mathcal{A} \times \mathcal{A}$ is a binary attack relation. An argument *a attacks* an argument *b* if $(a, b) \in \mathcal{R}_{att}$.

Dung's AF is adopted as the formal backbone of ArgRE because conflict resolution in this setting is inherently attack-centric. Proposals, critiques, and refinements can be uniformly represented as attack relations, without requiring richer internal argument structure. This abstraction is sufficient for the semantics computation in Phase 2, while remaining compatible with traceability-oriented extensions discussed later.

---

[1]Phase definitions are detailed in Section IV.

TABLE 1: Comparison of Multi-Agent RE Frameworks

| Framework | Agent Decomp. | Negotiation | Conflict Resolution | Argumentation | KAOS | Compliance |
|---|---|---|---|---|---|---|
| MARE [10] | By task | Single-turn | None | × | × | × |
| iReDev [11] | By knowledge | Human-in-loop | None | × | × | × |
| Elicitron [19] | By persona | None | None | × | × | × |
| QUARE [27]* | By quality | Dialectical | Heuristic (priority-weighted) | Informal | ✓ | ✓ |
| **ArgRE** | By quality | Dialectical | Argumentation semantics | Abstract argumentation | ✓ | ✓ |

*QUARE employs argumentation-inspired negotiation dynamics but does not construct formal argumentation frameworks or perform extension computation.

### 2) Argumentation Semantics

Given an AF, argumentation semantics define criteria for determining which sets of arguments (called extensions) are acceptable. We employ two well-established semantics: *grounded* and *preferred* (see Baroni et al. [40] for a survey, and Modgil and Caminada [41] for algorithms for computing extensions).

A set $S \subseteq \mathcal{A}$ is *conflict-free* if no two arguments in $S$ attack each other, i.e., $\nexists a, b \in S$ such that $(a, b) \in \mathcal{R}_{att}$. A conflict-free set $S$ *defends* an argument $a$ if, for every attacker $b \in \mathcal{A}$ of $a$, there exists $c \in S$ such that $(c, b) \in \mathcal{R}_{att}$. The set $S$ is *admissible* if it defends all its elements. Building on these notions, the *grounded extension* is the smallest complete extension, obtained as the least fixed point of the characteristic function:

$$F_{AF}(S) = \{a \in \mathcal{A} \mid S \text{ defends } a\}, \quad (2)$$

It is unique and represents the most conservative (skeptical) set of accepted arguments. A preferred extension is a maximal admissible set with respect to set inclusion; multiple preferred extensions may exist, representing alternative credulous acceptance positions.

*Intuitive summary.* Informally, Dung's framework accepts an argument if either (i) it is unattacked, or (ii) all its attackers are themselves counterattacked by accepted arguments. The grounded extension applies this criterion conservatively, admitting only arguments that are indisputably defended. Preferred extensions apply it more liberally, admitting maximal defensible sets. When the underlying attack graph is acyclic, both semantics yield the same extension.

### 3) Relevance to RE Negotiation

Dung's abstract argumentation is instantiated in the RE negotiation context as follows. An *argument* corresponds to a negotiation act: a requirement proposal, a critique of an existing proposal, or a refinement that revises a prior proposal. Formally, each argument is a tuple $a = \langle id, \tau, c, \alpha, q, \rho \rangle$, where $\tau \in \{proposal, critique, refinement\}$ denotes the act type. An *attack relation* $(a_i, a_j) \in \mathcal{R}_{att}$ holds when argument $a_i$ attacks argument $a_j$: a critique attacks the proposal it challenges, while a refinement attacks both the original proposal it supersedes and the critique it resolves.

Under this view, an *accepted extension* corresponds to the subset of requirements that survive negotiation under a given semantics. This mapping is practically significant because the choice of semantics determines how conservatively conflicts are resolved. Grounded semantics yields a skeptical baseline, admitting only uncontested or well-defended requirements, whereas preferred semantics permits larger admissible sets at the cost of potentially divergent interpretations.

*Illustrative example.*

Consider an autonomous-driving perception system. A Safety agent proposes argument $a_1$: "sensor fusion shall complete within 500 ms for dual-channel verification." An Efficiency agent raises argument $a_2$ (a critique): "500 ms exceeds the 30-ms planning-cycle budget; fusion must complete within 30 ms." Under Dung's framework, $a_2$ attacks $a_1$, i.e., $(a_2, a_1) \in \mathcal{R}_{att}$. The Safety agent then proposes a refinement $a_3$: "two-stage pipeline with fast path $\leq 30$ ms and thorough-path $\leq 500$ ms." This refinement *attacks* (supersedes) the original $a_1$ and also *attacks* (addresses) the critique $a_2$. Under grounded semantics, $a_3$ is accepted because its attackers (if any) are counterattacked by accepted arguments, yielding a traceable justification for relaxing the original 500-ms bound.

Under heuristic synthesis, by contrast, such constraints are typically merged into a single compromise requirement (e.g., "fusion shall complete within 100 ms") without preserving which agent's concern was prioritized or the rationale for the trade-off. In contrast, the argumentation graph renders this reasoning explicit and auditable. Section IV-H revisits this conflict in detail within the full ArgRE pipeline.

## IV. METHOD: ARGRE
### A. RESEARCH QUESTIONS

This study investigates four research questions:

**RQ1 (Interpretability):** Does argumentation-based resolution improve the transparency and traceability of conflict handling compared to heuristic resolution?

**RQ2 (Semantic Preservation):** Can ArgRE preserve semantic intent as effectively as heuristic conflict-resolution approaches?

**RQ3 (Requirement Quality):** How does ArgRE affect compliance coverage, verifiability, feasibility, and overall requirement quality?

**RQ4 (Structural Characterization):** What structural properties characterize the constructed argumentation graphs, and under what conditions do grounded and preferred extensions coincide or diverge?

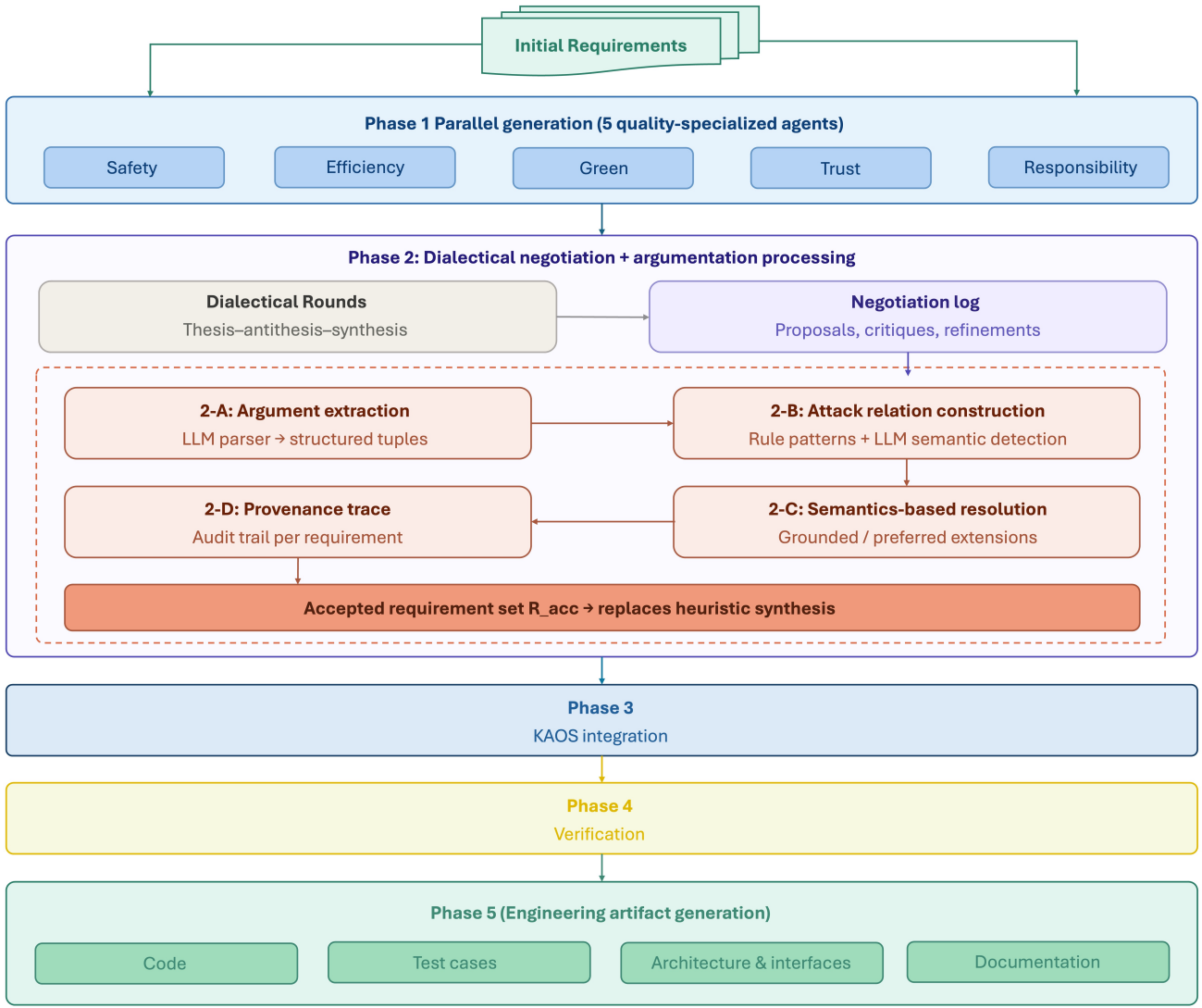

FIGURE 1: Overview of the ArgRE framework. A stakeholder provides the initial project requirements (top) and may intervene at three stages: inspecting the argumentation graph after Phase 2, overriding attack relations prior to re-solving, or reviewing flagged goals in Phase 4 (see Section VII-D). The pipeline comprises five phases: parallel generation (Phase 1), dialectical negotiation with formal argumentation (Phase 2), KAOS integration (Phase 3), verification (Phase 4), and artifact generation (Phase 5). The argumentation layer within Phase 2 (dashed box) constructs attack graphs, computes accepted sets under formal semantics, and records audit trails for accepted arguments.

### B. MULTI-AGENT ARCHITECTURE

*Pipeline Overview.*

ArgRE takes as input a natural-language project description specifying the system's purpose, domain, and key constraints (e.g., "Design a sensor-fusion module for a Level-4 autonomous vehicle compliant with ISO 26262 ASIL-D within a 30 ms planning cycle"), and produces three categories of output: (i) a verified KAOS goal model encoding negotiated requirements as a hierarchical goal graph, (ii) human-readable requirements documentation with traceability annotations linking each requirement to its argumentation provenance, and (iii) downstream engineering artifacts including test-case specifications and compliance evidence documents.

The pipeline follows a generate–negotiate–integrate–verify–deliver structure that aligns with RE workflows [1], [23] while introducing formal argumentation at the negotiation stage:

1) **Parallel Generation** elicits requirements independently from multiple quality perspectives independently, maximizing coverage and avoiding premature convergence.
2) **Dialectical Negotiation with Formal Argumentation** surfaces conflicts, resolves them under formal acceptability semantics, and records provenance traces.
3) **KAOS Integration** organizes the accepted requirement

set into a hierarchical goal structure with refinement links to support downstream verification and compliance mapping.
4) **Verification** checks structural compliance, identifying hallucinated or inconsistent requirements from the LLM backbone.
5) **Output Generation** produces auditable deliverables for external toolchains.

Phases 1, 3, 4, and 5 are adapted from the QUARE pipeline [27]. The principal architectural change lies in Phase 2, where ArgRE replaces heuristic, priority-weighted synthesis with Dung-style argumentation semantics.

*Design Rationale for Agent Decomposition.*

Task-based decomposition (as in MARE) assigns agents to pipeline stages, distributing workload but not inducing the structured disagreement required for meaningful negotiation. ArgRE instead adopts *quality-based decomposition*, in which each agent advocates for a distinct quality objective. This design deliberately introduces tension to surface trade-offs during dialectical rounds, while reducing prompt complexity per agent.

This decomposition is grounded in ISO/IEC 25010 [3] and extended with two additional concerns increasingly required in regulated contexts. Three agents map directly to ISO/IEC 25010: *Safety* (reliability and safety), *Efficiency* (performance efficiency), and *Trustworthiness* (security and privacy, including data-protection obligations under GDPR and ISO/IEC 27701). Two agents address concerns beyond the standard: *Green* (energy efficiency and resource footprint, aligned with ISO 14001) and *Responsibility* (ethical accountability and regulatory compliance, aligned with the EU AI Act and IEEE 7000). These five dimensions are selected as a core configuration for regulated RE settings, where safety, performance, security and privacy, sustainability, and accountability most directly influence acceptance decisions. Other ISO/IEC 25010 dimensions (e.g., usability, maintainability, portability) can be incorporated modularly as additional specialized agents when required by the project domain. Together with the Orchestrator, which manages pair scheduling, conflict state, and phase transitions, they constitute the six-role architecture of ArgRE. Each specialized agent operates using a prompt template comprising: (1) a role definition specifying the quality dimension and decision stance; (2) reasoning instructions for critique and refinement; and (3) an output schema enforcing JSON-structured requirements (goal ID, description, quality dimension, KAOS level, rationale).

### C. PHASE 1: PARALLEL GENERATION

In Phase 1, five quality-specialized agents process the same project description in parallel, each generating candidate requirements from its respective quality perspective. This fan-out design prioritizes coverage and diversity over early consensus.

TABLE 2: Specialized Agents in ArgRE

| Agent Role | Quality Concern | Core Responsibility |
|---|---|---|
| Safety | Safety & Reliability | Hazard identification and mitigation |
| Efficiency | Performance | Resource optimization and latency control |
| Green | Sustainability | Energy efficiency and carbon footprint |
| Trustworthiness | Security & Privacy | Data protection and access control |
| Responsibility | Ethical & Compliance | Regulatory and social responsibility |
| Orchestrator | System Integration | Multi-agent coordination and workflow |

Agents are functionally isolated through dedicated system prompts, preventing convergence to homogeneous outputs and preserving dimension-specific diversity. Each agent employs zero-shot prompting to ensure replicability across case studies. Outputs conform to a shared JSON schema (goal ID, description, quality dimension, KAOS level, rationale), enabling deterministic parsing and seamless downstream integration with negotiation and KAOS modeling.

The resulting requirement sets are intentionally redundant and potentially conflicting. ArgRE defers reconciliation to Phase 2, where conflicts are explicitly detected, debated, and resolved under formal argumentation semantics. For example, in the autonomous-driving case (Section IV-H), the Safety and Efficiency agents independently propose conflicting latency constraints that are subsequently resolved in Phase 2.

### D. PHASE 2: DIALECTICAL NEGOTIATION WITH FORMAL ARGUMENTATION

*Notation.*

To avoid ambiguity with the argument set $\mathcal{A}$, the set of specialized agents is denoted as $\mathcal{AG} = \{Ag_1, \ldots, Ag_n\}$, with individual agents denoted by $Ag_i$. Arguments remain denoted by $a_i$.

Phase 2 transforms potentially conflicting candidate requirements into an accepted set justified under argumentation semantics. It comprises four tightly coupled steps: negotiation protocol execution, argument and attack-graph construction, semantics-based resolution with provenance recording, and an optional cross-pair arbitration round for cyclic cross-quality conflicts. The default protocol is pairwise, as it localizes conflicts, improves interpretability, and maintains structural simplicity while still exposing key quality trade-offs. Cross-pair arbitration is introduced as an optional extension when the objective is to analyze cyclic cross-quality conflicts beyond this default setting.

1) Negotiation Protocol

*Conflict Detection.*

Before argument extraction, ArgRE performs a two-stage conflict detection process over candidate requirements generated in Phase 1 and updated during negotiation. In Stage 1, the Orchestrator computes pairwise cosine similarity using

`bert-base-uncased` embeddings and flags requirement pairs exceeding a threshold $\tau = 0.85$.

In Stage 2, an LLM-based classifier analyzes each flagged pair and assigns one of three labels: (a) *redundant*, (b) *resource-bound conflict*, or (c) *logical incompatibility*. Redundant pairs are candidates for consolidation, whereas conflict labels trigger dialectical debate and subsequent argumentation processing. Resource-bound conflicts arise when quality objectives compete for limited resources (e.g., latency budget, compute, or energy), whereas logical incompatibilities arise when requirements impose mutually exclusive system states. This distinction is retained in negotiation metadata and later reflected in attack-graph rationales.

In the autonomous-driving case, the Safety agent's 500-ms proposal and the Efficiency agent's 30-ms constraint are classified as a resource-bound conflict in Stage 2, thereby triggering dialectical debate in subsequent rounds.

*Dialectical Rounds.*

Negotiation proceeds in bounded, multi-round thesis–antithesis–synthesis cycles. Let $\mathcal{AG} = \{Ag_1, \ldots, Ag_n\}$ denote the set of specialized agents with quality objectives $Q_i$ and priority weights $w_i$. Conceptually, the negotiation seeks a requirement formulation $r^*$ that maximizes the weighted objective $\sum_{i=1}^{n} w_i \cdot Q_i(r)$. ArgRE does not optimize this objective directly; instead, the argumentation layer captures trade-offs through attack and defense dynamics, with priority weights $w_{q(a)}$ applied only during preferred-extension selection (Section IV-D3).

At round $k$, a focus agent proposes or refines requirement $r_k$, peer agents submit critiques, and the Orchestrator synthesizes the next candidate:

$$r_{k+1} = G(r_k, \text{critique}(Ag_j, r_k)), \quad (3)$$

where $G(\cdot)$ is implemented as an LLM call that takes the current requirement and all peer critiques as input and produces a revised formulation addressing the identified concerns. Scheduling follows a round-robin scheme, ensuring that each specialized agent serves as the focus agent exactly once per round.

Termination occurs when either (i) the BERTScore similarity between successive rounds exceeds $1 - \epsilon$ ($\epsilon = 0.02$), indicating convergence, or (ii) the round cap $N = 3$ is reached. In practice, condition (ii) is typically binding, while condition (i) serves as an early-exit mechanism for low-conflict cases that stabilize before reaching the cap.

2) Argument and Attack-graph Construction

The negotiation protocol produces structured interaction logs containing proposals, critiques, and refinements. To make conflict rationales explicit and enable formal extension computation, ArgRE transforms these logs into an AF $\langle \mathcal{A}, \mathcal{R}_{att} \rangle$, whose nodes correspond to extracted arguments and whose directed edges represent labeled attack relations. The resulting graph serves both as (i) input to semantics-based resolution and (ii) an auditable explanation layer that records which requirements conflict, why, and how those conflicts are resolved.

*Argument Extraction.*

The first step is to extract structured arguments from the negotiation artifacts generated during the dialectical rounds. An argument is defined as the tuple:

$$a = \langle id, \tau, c, \alpha, q, \rho \rangle, \quad (4)$$

where *id* is a unique identifier; $\tau \in \{\text{proposal, critique, refinement}\}$ denotes the argument type; $c$ represents the requirement-related content; $\alpha \in \{Ag_1, \ldots, Ag_5\}$ identifies the originating specialized agent; $q \in \{q_1, \ldots, q_5\}$ denotes the associated quality dimension; and $\rho$ encodes the rationale or justification chain.

Argument extraction is performed by an LLM-based parser that processes the negotiation log and identifies three categories of argumentative acts: (1) **Proposals**, i.e., new or revised requirement statements introduced by a focus agent; (2) **Critiques**, i.e., explicit objections raised by peer agents, typically citing constraint violations or quality trade-offs; and (3) **Refinements**, i.e., modified proposals that address critiques while preserving core intent. The parser outputs a set of arguments $\mathcal{A} = \{a_1, a_2, \ldots, a_m\}$, with trace links to the source turn, originating agent, and associated quality dimension.

*Attack Relation Construction.*

Given the extracted argument set $\mathcal{A}$, ArgRE constructs the attack relation $\mathcal{R}_{att}$ using a modular approach that combines three deterministic structural rules with a threshold-gated LLM-based pathway for cross-pair semantic conflicts. The structural rules provide complete intra-pair coverage under the pairwise dialectical protocol, while the LLM pathway captures semantic conflicts not covered by these patterns. In the main experiment, the effective threshold is set to $\theta_{\text{eff}} = 0.85$ to prioritize precision.

Throughout this paper, $(a_i, a_j) \in \mathcal{R}_{att}$ denotes that argument $a_i$ *attacks* argument $a_j$. Equivalently, $a_i \rightarrow a_j$ indicates the direction of attack from the attacker to the target.

**Rule-based Attack Patterns.** We define three structural attack patterns, identified deterministically from negotiation logs:

**Pattern 1 (Critique → Proposal):** If argument $a_i$ is a critique that explicitly references and rejects proposal $a_j$, then $(a_i, a_j) \in \mathcal{R}_{att}$.

**Pattern 2 (Refinement → Original):** If argument $a_i$ is a refinement that supersedes an earlier proposal $a_j$ from the same agent, then $(a_i, a_j) \in \mathcal{R}_{att}$.

**Pattern 3 (Refinement → Critique):** If argument $a_i$ is a refinement that resolves the concerns raised by critique $a_j$, then $(a_i, a_j) \in \mathcal{R}_{att}$. This edge represents a counter-attack: the refinement defends its parent proposal by neutralizing the objection.

Section IV-H and Figure 3 illustrate all three patterns in the autonomous-driving conflict.

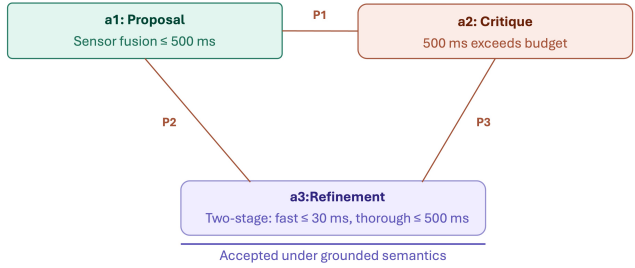

FIGURE 2: Basic attack pattern structure. P1: critique attacks a proposal (Pattern 1). P2: a refinement supersedes the original proposal (Pattern 2). P3: a refinement addresses a critique (Pattern 3). Under grounded semantics, $a_3$ is accepted as it defeats both the original proposal and the critique.

*Theoretical grounding.*

These three patterns instantiate standard argumentation moves from the defeasible reasoning literature. Pattern 1 corresponds to a counterargument, in which a critique directly challenges the conclusion of a proposal. Pattern 2 corresponds to proper defeat, where a more specific refinement renders an earlier proposal obsolete. Pattern 3 corresponds to a defeater of a defeater (reinstatement), in which a refinement neutralizes a critique and restores the original line of reasoning.

All three moves are formalized in the dialectical trees of Defeasible Logic Programming [42], where arguments, counterarguments, and reinstating defeaters interact through this layered structure. The contribution here lies not in the attack typology itself, but in its operationalization within multi-agent RE negotiation logs, where proposals, critiques, and refinements provide deterministic structural cues that map directly to these established categories.

**LLM-based Semantic Conflict Detection (Threshold-gated).** Patterns 1–3 generate attacks only *within* a single pairwise negotiation session, as the structural cues they rely on are absent across sessions. Cross-session semantic conflicts, where requirements from independent negotiations impose incompatible system states or compete for shared resources, require natural-language understanding. Recent work shows that NLI-based classifiers can detect requirement-level contradictions [43], and that LLM-based approaches can identify requirement conflicts with high precision [44]. Building on these findings, ArgRE introduces a threshold-gated LLM pathway for cross-session conflicts.

From each completed session, ArgRE selects surviving arguments: proposal and refinement arguments not defeated by intra-session rule-based attacks (if all candidates are defeated, the latest refinement is retained as the session representative). Every unordered pair of survivors drawn from *distinct* sessions is then evaluated by an LLM classifier, which determines whether realizing one argument renders the other infeasible or materially more difficult—capturing incompatible global-state constraints, quantitative resource conflicts, and mutually exclusive architectural commitments.

An attack edge is added to $\mathcal{R}_{att}$ only when the classifier reports a conflict with confidence $c \geq \theta_{\text{eff}}$, where $\theta_{\text{eff}} = \max(\theta_{\text{floor}}, \theta)$ with nominal $\theta = 0.7$ and $\theta_{\text{floor}} = 0.85$ (Section V-C), yielding $\theta_{\text{eff}} = 0.85$ in all reported runs. At this conservative setting, no LLM-derived edges are retained ($\mathcal{R}_{\text{llm}} = \emptyset$), and all baseline attacks are explained by deterministic structural patterns. Section VI-D3 varies $\theta_{\text{eff}}$ and shows that lowering it progressively activates cross-session classifications, increasing GCI and inducing semantics divergence. The architecture also supports recording support relations $\mathcal{R}_{sup} = \{(a_i, a_j) \mid a_i \text{ supports } a_j\}$, although these are not required for extension computation in the main experiments. When recorded, support is operationalized as a defense pattern within Dung's standard AF: if $a_i$ supports $a_j$ and $a_k$ attacks $a_j$, then $a_i$ is expected to attack $a_k$. This preserves compatibility with Dung's framework while capturing supportive interactions for traceability.

### 3) Semantics-based Resolution and Traceability

Given the constructed argumentation framework $AF = \langle \mathcal{A}, \mathcal{R}_{att} \rangle$, ArgRE computes accepted arguments under two standard semantics. The grounded extension $\mathcal{E}_g$ is computed as the least fixed point of the characteristic function $F_{AF}$:

$$\mathcal{E}_g = \text{lfp}(F_{AF}) = F_{AF}^0(\emptyset) \cup F_{AF}^1(\emptyset) \cup \cdots \quad (5)$$

This yields the most skeptical accepted set: only arguments that are unattacked or defended by already accepted arguments are included, corresponding, in RE terms, to requirements that face no unresolved opposition.

Preferred extensions $\{\mathcal{E}_p^1, \mathcal{E}_p^2, \ldots\}$ are maximal admissible sets. When multiple preferred extensions exist, two resolution strategies are applied: (1) *intersection*, $\mathcal{E}_p^\cap = \bigcap_i \mathcal{E}_p^i$, retaining only arguments present in all extensions; and (2) *priority-guided selection*, which selects the extension that maximizes $\sum_{a \in \mathcal{E}_p^i} w_{q(a)}$, where $w_{q(a)}$ denotes the project-specific priority weight associated with the quality dimension of argument $a$. The accepted requirements are then derived from the chosen extension:

$$\mathcal{R}^{acc} = \{ r(a) \mid a \in \mathcal{E}^*(AF) \land \tau(a) \in \{\text{proposal}, \text{refinement}\} \}, \quad (6)$$

where $\mathcal{E}^*$ denotes the selected extension (grounded or preferred) and $r(a)$ extracts the requirement content from argument $a$. Critiques are excluded because they represent objections rather than actionable requirements.

For example, in the autonomous-driving case, the final three-component architecture ($a_5$) and the consensus acceptance ($a_6$) are unattacked arguments that enter the grounded extension directly, while superseded intermediate proposals ($a_1$, $a_3$) and resolved critiques ($a_2$, $a_4$) are excluded from $\mathcal{R}^{acc}$.

*Provenance and Traceability.*

A key output of this resolution step is a traceable record for each accepted requirement. For each $r \in \mathcal{R}^{acc}$, ArgRE

records the originating argument *a* together with its type, agent, quality dimension, all attack relations involving *a*, the chain of counter-attacks through which acceptance is established (hereafter, *defense chain*), and the specific semantics under which it was accepted. This information is stored as structured metadata and can be visualized as an argumentation graph, enabling stakeholders to audit the rationale behind each accepted requirement.

4) Cross-Pair Arbitration Round

The pairwise dialectical protocol described above localizes conflicts within agent pairs and therefore produces acyclic graphs (GCI = 0). To investigate ArgRE's behavior under cross-quality conflicts, we introduce an optional *cross-pair arbitration round* that runs once after the standard pairwise rounds complete.

The Orchestrator first identifies shared-resource overlaps: requirement pairs from different pairwise negotiations that reference the same finite resource (e.g., compute budget or latency headroom), reusing the Stage-1 cosine-similarity filter ($\tau = 0.85$) across negotiation chains. For each flagged overlap, the two originating agents enter a single additional thesis–antithesis exchange: agent $Ag_i$ critiques the accepted proposal of $Ag_j$ from its own quality perspective, and vice versa. Because both critiques target each other's proposals, the resulting Pattern 1 attacks are bidirectional, creating mutual attacks $(a_x, a_y), (a_y, a_x) \in \mathcal{R}_{\text{att}}$ that introduce cycles. In practice, 1–2 overlaps are detected per case study, yielding GCI values in the range $[0.20, 0.29]$, where GCI denotes the proportion of arguments in cyclic strongly connected components. The AF solver then computes grounded and preferred extensions over the augmented graph, producing distinct accepted sets when cycles are present. The same threshold setting $\theta_{\text{eff}} = 0.85$ is retained throughout so that any change in GCI is attributable solely to the protocol extension.

E. PHASE 3: KAOS INTEGRATION

The accepted requirement set $\mathcal{R}^{acc}$ from Phase 2 is transformed into a coherent KAOS goal structure through semantic deduplication, cross-agent parent–child stitching, and DAG topology validation. Integration preserves negotiated intent while enforcing structural constraints needed for downstream verification.

Semantic deduplication merges near-equivalent requirements with overlapping intent while retaining trace links to source arguments. Cross-agent stitching then connects goals generated by different quality agents into a unified hierarchy, resolving fragmentation that naturally arises from quality-specialized generation.

ArgRE enforces a three-level KAOS hierarchy (Strategic, Tactical, Operational) and validates the resulting structure as a directed acyclic graph. Cycles are prohibited because circular goal dependencies undermine traceability and compliance reasoning. When structural violations occur, correction rules rewire topology while preserving the semantic commitments established in Phase 2.

F. PHASE 4: VERIFICATION

ArgRE applies three-layer verification over the integrated KAOS model. Each layer operates as a pure validation pass: it may flag issues and attach metadata but does not modify the goal content established in Phase 3.

*a: Layer 1: Structural Rule Checking.*

A deterministic rule engine checks the following properties over the KAOS goal graph $G = (V, E)$, where $V$ is the set of goals and $E$ the set of refinement links:

1) **Schema completeness:** Every goal $v \in V$ must include all mandatory fields: goal ID, description, quality dimension, KAOS level $\in \{Strategic, Tactical, Operational\}$, and rationale.
2) **DAG validity:** The goal graph must be a directed acyclic graph. Cycles are detected via topological sort; any cycle triggers a correction rule that rewires the lowest-priority edge.
3) **Refinement completeness:** Each non-leaf goal must have at least one child via AND/OR refinement. Leaf goals must reside at the Operational level.
4) **Root connectivity:** Every goal must be reachable from at least one Strategic-level root.
5) **Cross-reference integrity:** All argument IDs referenced in provenance annotations must exist in the Phase 2 argumentation graph.

Violations are logged with severity (error or warning). Error-level issues (e.g., cycles, missing mandatory fields) block progression to Layer 2, whereas warning-level issues (e.g., single-child AND refinements) are flagged for review.

*b: Layer 2: Retrieval-augmented Hallucination Detection.*

Each goal description is embedded using `text-embedding-ada-002` and queried against a ChromaDB-backed corpus containing relevant standards (ISO 26262, ISO 27001, ISO/IEC 25010) and domain-specific technical references. A goal is flagged as a potential hallucination if its nearest-neighbor cosine similarity falls below $\tau_h = 0.60$, indicating that no corpus passage substantively supports the stated requirement. Flagged goals are annotated with the closest corpus passage and a confidence score for downstream human review. This layer targets LLM-generated content that is linguistically plausible but technically ungrounded.

*c: Layer 3: Standards-oriented Compliance Verification.*

For each applicable regulatory clause (e.g., ISO 26262 Part 3 clauses for the AD case, ISO 27001 Annex A controls for financial cases), an LLM-based entailment classifier determines whether at least one accepted requirement in $\mathcal{R}^{\text{acc}}$ satisfies the clause. The classifier receives the clause text and all candidate requirements as input and outputs a binary entailment judgment with a rationale. Compliance coverage is computed as the proportion of applicable clauses for which at least one entailing requirement exists. Algorithm 1 summarizes the three-layer procedure; Layer 1 aggregates the

**Algorithm 1** Three-Layer Verification (Phase 4)

**Require:** KAOS goal graph $G = (V, E)$, argumentation graph $G_{AF}$, accepted set $\mathcal{R}^{acc}$, standards corpus $\mathcal{C}$, threshold $\tau_h$
**Ensure:** Annotated $G$, compliance coverage $\gamma$
1: **Layer 1:** $S \leftarrow \text{StructuralCheck}(G, E, G_{AF})$ ▷ schema, DAG, refinement, roots, arg-ID integrity
2: **if** $S$ contains error-level violation **then**
3: 
4:    **return** failure ▷ block Layers 2–3
5: **end if**
6: **Layer 2:**
7: **for** each $v \in V$ **do**
8:    $e_v \leftarrow \text{Embed}(\text{description}(v))$ ▷ `text-embedding-ada-002`
9:    $(s^*, \sigma) \leftarrow \text{NearestNeighbor}(e_v, \mathcal{C})$ ▷ ChromaDB; cosine $\sigma$
10:    **if** $\sigma < \tau_h$ **then**
11:      annotate $v$ with $(s^*, \sigma)$
12:    **end if**
13: **end for**
14: **Layer 3:** $\mathcal{K} \leftarrow \text{ApplicableClauses}(\mathcal{C})$; $sat \leftarrow 0$
15: **for** each clause $k \in \mathcal{K}$ **do**
16:    **if** $\exists r \in \mathcal{R}^{acc} : \text{LLMEntails}(k, r)$ **then**
17:      $sat \leftarrow sat + 1$
18:    **end if**
19: **end for**
20: $\gamma \leftarrow |\mathcal{K}|^{-1} \cdot sat$
21: **return** annotated $G, \gamma$

**Algorithm 2** Argumentation-based Negotiation Resolution

**Require:** Negotiation log $\mathcal{L}$ from dialectical rounds, semantics choice $\sigma \in \{grounded, preferred\}$
**Ensure:** Accepted requirement set $\mathcal{R}^{acc}$, argumentation graph $G_{AF}$
1: **Step 1: Argument Extraction**
2: $\mathcal{A} \leftarrow \text{ExtractArguments}(\mathcal{L})$ ▷ Section IV-D2
3: **Step 2: Attack Relation Construction**
4: $\mathcal{R}_{rule} \leftarrow \text{RuleBasedAttacks}(\mathcal{A})$ ▷ Patterns 1–3
5: $\mathcal{R}_{llm} \leftarrow \text{LLMConflictDetection}(\mathcal{A}, \theta)$ ▷ Cross-pair semantic conflicts (threshold-gated; § VI-D3)
6: $\mathcal{R}_{att} \leftarrow \mathcal{R}_{rule} \cup \mathcal{R}_{llm}$
7: **Step 3: Semantics-Based Resolution**
8: $AF \leftarrow \langle \mathcal{A}, \mathcal{R}_{att} \rangle$
9: **if** $\sigma = grounded$ **then**
10:    $E^* \leftarrow \text{GroundedExtension}(AF)$
11: **else**
12:    $\{E_p^1, \ldots\} \leftarrow \text{PreferredExtensions}(AF)$
13:    $E^* \leftarrow \text{SelectExtension}(\{E_p^i\}, \mathbf{w})$
14: **end if**
15: **Step 4: Requirement Derivation**
16: $\mathcal{R}^{acc} \leftarrow \{r(a) \mid a \in E^*, \tau(a) \neq \text{critique}\}$
17: $G_{AF} \leftarrow (\mathcal{A}, \mathcal{R}_{att}, E^*)$ ▷ Provenance graph
18: **return** $\mathcal{R}^{acc}, G_{AF}$

structural rules of the preceding paragraphs into a single StructuralCheck routine.

*d: Preservation of negotiated intent.*

Empirically, BERTScore F1 between Phase 3 and Phase 4 outputs remains effectively unchanged across all case studies (mean $\Delta < 0.1$ percentage points), confirming that the verification layer annotates and flags issues without altering the underlying requirement content.

### G. PHASE 5: OUTPUT GENERATION

ArgRE finalizes standardized artifacts by mapping verified KAOS goals and associated verification metadata into multiple deliverables. Core outputs include KAOS models (JSON and GSN XML), human-readable requirements documentation, and downstream engineering artifacts such as test-case specifications and architecture-support documents.

The generation engine is template-driven: it preserves trace links to argument IDs, attack relations, and verification evidence while exporting format-specific artifacts for analysis, auditing, and integration into external toolchains. Algorithm 2 summarizes the argumentation-based resolution subprocedure within Phase 2 after the dialectical rounds have produced the negotiation log.

### H. RUNNING EXAMPLE: SENSOR FUSION LATENCY CONFLICT

To illustrate ArgRE's argumentation layer, we trace a representative conflict through the full pipeline. The example derives from one of our five evaluation case studies: an *autonomous-driving perception system* with the following project description:

> "Design the perception subsystem for a Level-4 autonomous vehicle. The system must comply with ISO 26262 ASIL-D safety requirements, operate within a 30 ms planning-cycle budget, and minimize energy consumption for battery-electric deployment."

This description serves as the sole input to Phase 1. The real-time planning constraint is instantiated from a project-specific Apollo design specification (Baidu Apollo open-source autonomous-driving platform) used in the evaluation scenario. The core conflict emerges because the Safety agent advocates a conservative dual-channel sensor-fusion verification budget of 500 ms, whereas the Efficiency agent requires fusion to complete within the 30-ms planning cycle to meet real-time constraints. This tension reflects a safety–latency trade-off rather than a direct standard violation. We next trace how this conflict propagates through argument extraction, attack-graph construction, and semantics-based resolution.

*1) Negotiation Artifacts*

During Phase 2 dialectical rounds, the following negotiation exchanges are recorded:

- **Round 1:** The Safety agent proposes a sensor-fusion latency of $\leq 500$ ms for conservative dual-channel verifica-

tion. The Efficiency agent critiques this proposal: "500 ms exceeds the available planning-cycle budget; fusion must complete within 30 ms under the real-time constraint." Status: unresolved.

- **Round 2:** The Safety agent refines the proposal into a two-stage pipeline: fast-path $\leq$ 30 ms (for 95% of cases) and thorough-path $\leq$ 500 ms (for anomaly-triggered cases). The Efficiency agent partially accepts but stipulates additional constraints: thorough-path must be asynchronous, fast-path success rate $\geq$ 99%, and fallback mechanisms ensure planning continuity. Status: partial.
- **Round 3:** The Safety agent proposes the final three-component architecture. The Efficiency agent confirms consensus. Status: resolved.

#### 2) Argument Extraction

The parser extracts the following structured arguments from the negotiation log, capturing type, originating agent, and content:
- $a_1$: [proposal, Safety, "Sensor fusion $\leq$ 500 ms for conservative dual-channel verification"]
- $a_2$: [critique, Efficiency, "500 ms exceeds budget; require $\leq$ 30 ms"]
- $a_3$: [refinement, Safety, "Two-stage pipeline: fast $\leq$ 30 ms (95% cases), thorough $\leq$ 500 ms (anomaly-triggered)"]
- $a_4$: [critique, Efficiency, "Thorough-path must be asynchronous; fast-path $\geq$ 99%; fallback ensures planning continuity"]
- $a_5$: [refinement, Safety, "Final three-component architecture: fast-path sync $\leq$ 30 ms $\geq$ 99%, thorough-path async $\leq$ 500 ms, fallback included"]
- $a_6$: [proposal, Efficiency, "Accept three-component architecture"]

#### 3) Attack Graph

Applying the rule-based and semantic attack patterns yields the following edges:
- $(a_2, a_1)$: Critique $\rightarrow$ Proposal (Pattern 1)
- $(a_3, a_1)$: Refinement supersedes original proposal (Pattern 2)
- $(a_3, a_2)$: Refinement addresses critique (Pattern 3)
- $(a_4, a_3)$: Critique $\rightarrow$ Refinement (Pattern 1)
- $(a_5, a_3)$: Further refinement supersedes prior refinement (Pattern 2)
- $(a_5, a_4)$: Refinement addresses critique (Pattern 3)
- $(a_6, a_2)$: Consensus proposal attacks original critique (semantic/LLM-based)

Figure 3 depicts the resulting argumentation graph. From top to bottom, the negotiation unfolds as follows: i) Initial conflict: The Safety agent's proposal $a_1$ and the Efficiency agent's critique $a_2$ are in direct opposition, connected by a P1 attack edge. ii) First refinement cycle: Argument $a_3$ supersedes $a_1$ (P2) and counters $a_2$ (P3), while $a_4$ introduces further constraints on $a_3$ (P1). iii) Final resolution: Argument $a_5$ supersedes $a_3$ (P2) and addresses $a_4$ (P3), and the consensus proposal $a_6$ confirms acceptance.

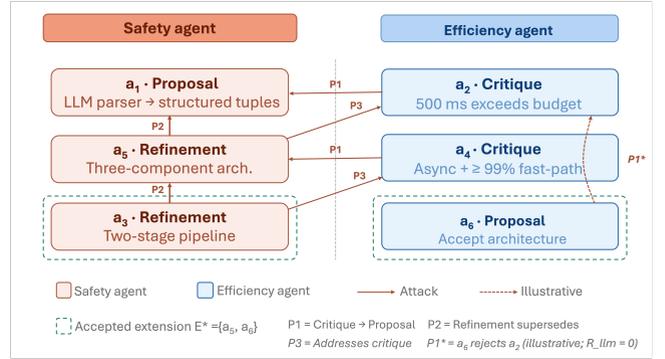

FIGURE 3: ArgRE argumentation graph for the autonomous-driving running example (Section IV-H). Each node represents an argument extracted from the negotiation log, with its type (proposal, critique, or refinement) and a brief content summary. Orange nodes originate from the Safety agent, blue nodes from the Efficiency agent. Red directed edges indicate attacks, labeled by pattern: P1 = critique objects to a proposal, P2 = refinement supersedes an earlier version, P3 = refinement resolves a critique. The dashed edge from $a_6$ to $a_2$ denotes a semantic attack, where the consensus acceptance invalidates the original critique. The bottom of the graph highlights the accepted extension $\mathcal{E}^* = \{a_5, a_6\}$, which is the same under both grounded and preferred semantics.

Figure 4 illustrates how the accepted extension from the argumentation outcome is integrated into a KAOS goal hierarchy in Phase 3, linking the negotiated three-component architecture to Tactical and Operational goals for downstream verification, standards compliance, and traceable engineering artifacts.

#### 4) Accepted Extension

Under grounded semantics, the unattacked arguments $a_5$ and $a_6$ are directly accepted, while $a_4$, being attacked by $a_5$ without defense, is rejected. Tracing backward, the grounded extension is therefore $\mathcal{E}_g = \{a_1, a_5, a_6\}$.

The co-acceptance of $a_1$ and $a_5$ can be explained via Dung's semantics. Argument $a_1$ is attacked by $a_2$ (Pattern 1) and $a_3$ (Pattern 2). Critique $a_2$ is itself attacked by $a_3$ (Pattern 3) and by $a_6$ (semantic edge). Since $a_6 \in \mathcal{E}_g$, $a_2$ is rejected. Refinement $a_3$ is attacked by $a_4$ and $a_5$ (Patterns 1 and 2). With $a_5 \in \mathcal{E}_g$, $a_3$ is rejected. With both attackers of $a_1$ absent from the grounded extension, $a_1$ is thus acceptable alongside $a_5$ and $a_6$. From an RE perspective, $a_5$ represents the *operational requirement* specifying the implemented architecture, while $a_1$ captures the *original quality intent* that motivated the refinement chain. During Phase 3 (KAOS integration), semantic deduplication merges such subsumption pairs, retaining $a_5$ as the operational goal and preserving $a_1$ as a traceable ancestor. This approach ensures the argumentation layer maintains a complete audit trail from the final requirement back to the initial quality objective. Under preferred semantics, the unique preferred extension is likewise $\{a_1, a_5, a_6\}$, as the acyclic graph admits no alternative admissible sets. The audit trail indicates that the accepted requirement $a_5$ originated from the SafetyAgent, was shaped by critiques from EfficiencyAgent ($a_2$ and $a_4$), and achieved consensus through iterative refinements. All steps are formally traceable within the argumentation graph.

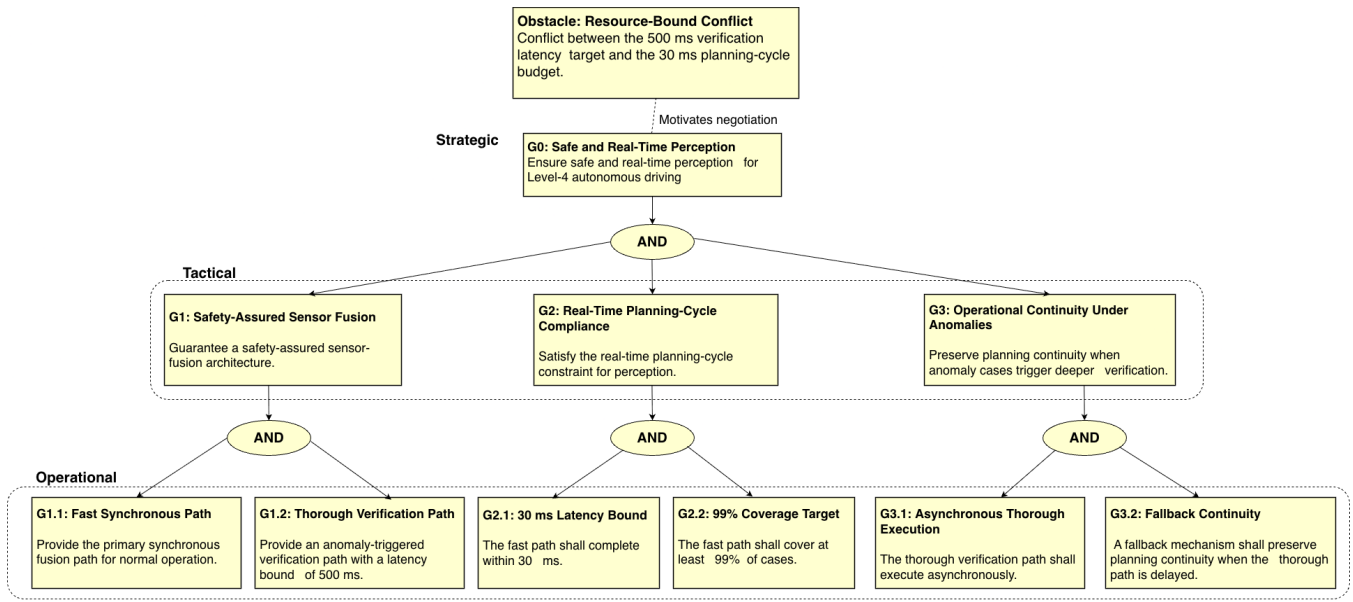

FIGURE 4: KAOS goal model for the autonomous-driving running example. The accepted requirements are organized into a three-level hierarchy: Strategic, Tactical, and Operational goals, connected via AND-refinement links. The Strategic goal is refined into three Tactical concerns: safety-assured sensor fusion, real-time planning-cycle compliance, and operational continuity under anomalies. These, in turn, are refined into Operational goals representing the 30-ms fast path, 99% coverage target, asynchronous thorough verification, and fallback continuity, reflecting the negotiated outcomes from the argumentation layer.

5) Provenance Trace Walkthrough

To illustrate ArgRE's traceability, we provide a complete backward trace for the accepted requirement $a_5$ from the running example.

**Trace Card 1: Accepted Requirement and Context**

Requirement: "Three-component architecture: fast synchronous path $\leq$ 30 ms with $\geq$ 99% coverage, thorough asynchronous path $\leq$ 500 ms for anomaly cases, and fallback to preserve planning continuity."
Origin: $a_5$, refinement, SafetyAgent, round 3.
Accepted under: grounded and preferred semantics ($a_5$ is unattacked).

**Trace Card 2: Backward Trace (Counter-attack Chain)**

The backward trace links $a_5$ through two refinement edges (P2) and two critique-resolution edges (P3) back to the original proposal $a_1$, spanning Safety ($a_1, a_3, a_5$) and Efficiency ($a_2, a_4, a_6$) dimensions over a resource-bound latency conflict. Each intermediate step is recorded with its pattern label, enabling stakeholders to audit the full negotiation history.

TABLE 3: Summary of Case Studies

| Case Study | Domain | Core Focus | Standard |
|---|---|---|---|
| Autonomous Driving | Automotive | Safety | ISO 26262 |
| ATM System | Finance | Security | ISO 27001 |
| Library System | Management | Scalability | ISO 27001 |
| RollCall System | Management | Functional | ISO 27001 |
| Bookkeeping System | Finance | Data Integrity | ISO 27001 |

## V. EXPERIMENTAL SETUP

### A. CASE STUDIES

We evaluate five case studies spanning safety-critical, financial, and information-system domains (Table 3).

### B. COMPARED METHODS

We compare five configurations: (1)MARE, a task-specialized multi-agent framework reimplemented in OpenReBench; (2)iReDev, a knowledge-driven multi-agent framework reimplemented in OpenReBench; (3)ArgRE-NoAF, an ablation variant that replaces formal argumentation with priority-weighted heuristic synthesis; (4)ArgRE (Grounded), the full system with grounded semantics; and (5)ArgRE (Preferred), the full system with preferred semantics and priority-guided extension selection.

### C. IMPLEMENTATION DETAILS

All experiments use `gpt-4o-mini-2024-07-18` as the LLM backbone with temperature fixed at 0.7 (except for attack detection, where it is set to 1.0 to encourage a broader distribution of confidence scores for improved threshold discrimination). The nominal confidence threshold for LLM-based attack detection is $\theta = 0.7$; the implementation applies $\theta_{\text{eff}} = \max(\theta_{\text{floor}}, \theta)$ with $\theta_{\text{floor}} = 0.85$, yielding $\theta_{\text{eff}} = 0.85$ and reducing spurious cross-pair attacks. OpenReBench exposes $\theta_{\text{floor}}$ as a configurable hyperparameter, which is kept fixed across all reported runs. Section VI-D3 reports a sensitivity analysis sweeping $\theta_{\text{eff}} \in \{0.50, 0.60, 0.70, 0.80, 0.85\}$ to characterize how the LLM pathway's contribution varies with threshold configuration. The argumentation layer is implemented as a custom Python module within OpenReBench, and semantics are computed using an implementation of Dung's algorithms. Each configuration is executed three times with different random seeds (101, 202, 303), and results are averaged under identical infrastructure conditions. To evaluate ArgRE under cross-quality conflicts in which grounded and preferred semantics may diverge, we additionally introduce a *Complex Conflict* scenario for each case study. This scenario extends the basic pairwise protocol with the cross-pair arbitration round described in Section IV-D4. All other parameters are kept identical ($\theta_{\text{eff}} = 0.85$, identical seeds, and the same backbone model), ensuring that observed differences are attributable solely to the protocol extension. Table 4 reports both scenarios side by side.

Priority weights for preferred-extension selection are set to uniform values ($w_q = 0.2$). For the safety-critical AD case, Safety is assigned an increased weight of $w_q = 0.3$, while the remaining four dimensions are set to $0.175$ each. These weights affect outcomes only when multiple preferred extensions exist, which occurs in the Complex Conflict scenario (GCI > 0) but not under the pairwise protocol (GCI = 0), where grounded and preferred extensions coincide.

### D. EVALUATION METRICS

The metrics are designed to answer two questions that a practitioner considering ArgRE would typically ask:

1) **Does the argumentation layer preserve requirement quality?** Introducing a formal argumentation layer may distort or filter useful requirements. We evaluate this risk using established RE metrics, including BERTScore for semantic preservation, compliance coverage for standards alignment, and ISO/IEC/IEEE 29148 quality scores for Verifiability and Feasibility. A desirable outcome is parity with the heuristic baseline on these measures.
2) **Does the argumentation layer *add* auditable decision rationale?** Heuristic synthesis produces final requirements but lacks structured explanations of *why* requirements are retained, revised, or rejected, a limitation often raised in ISO 26262 and EU AI Act Article 11 audits. We therefore introduce three metrics to quantify this added value: TC for provenance coverage, DJS for human-assessed rationale quality, and GCI as a structural indicator of when semantic selection becomes operationally relevant.

Metrics in group (1) are drawn from established RE evaluation practice, whereas metrics in group (2) are introduced in this work to evaluate properties specific to argumentation-based resolution. The two groups reflect distinct practitioner concerns in regulated RE settings. Group (1) metrics address the primary adoption risk: that introducing a formal argumentation layer may degrade requirement quality or reduce coverage of applicable standards. Group (2) metrics address the primary value proposition: that the argumentation layer should produce auditable decision rationale that satisfies traceability requirements in standards such as ISO 26262 and Article 11 of the EU AI Act, a capability absent in heuristic-synthesis baselines. We evaluate both groups to test whether the argumentation layer functions as a transparent overlay, adding traceability without degrading requirement quality.

#### 1) Core Metrics

We adopt the following metrics. Requirement Count is the total number of unique generated requirements. Semantic Preservation measures set-level similarity between Phase 1 and Phase 3 outputs using BERTScore [45]: we build a pairwise score matrix, apply optimal bipartite matching to maximize total alignment, and average matched scores. Conflict Resolution Rate (CRR) is the proportion of detected conflicts that are resolved. Compliance Coverage is the proportion of applicable regulatory clauses satisfied, measured via RAG-based clause entailment. Verifiability and Feasibility are ISO/IEC/IEEE 29148 quality scores on a 1–5 scale. Coverage Uniformity (CU) and Minimum Axis Coverage (MAC) capture per-axis distribution balance.

#### 2) Metrics Introduced in This Work

We introduce three metrics to evaluate the argumentation layer.

TC measures the proportion of accepted requirements for which a complete provenance trace, from accepted argument back to originating proposal through all intermediate critiques and refinements, can be reconstructed from the argumentation graph:

$$TC = \frac{|\{r \in \mathcal{R}^{acc} \mid \text{trace}(r) \text{ is complete}\}|}{|\mathcal{R}^{acc}|} \quad (7)$$

DJS is a human-evaluated score on a 1–5 scale measuring whether the artifacts supporting conflict resolution provide an explicit rationale for each accepted requirement. DJS is evaluated on three case studies (AD, ATM, Bookkeeping) in a single-blind, between-methods comparison. Three raters (R1–R3), none affiliated with this work, assessed each accepted requirement under two conditions: ArgRE with the argumentation graph and trace cards, and ArgRE-NoAF with only the final requirement list. The two conditions were labeled "Method A" and "Method B" under randomized assignment. Raters were not informed which method corresponded to the proposed approach. All raters completed calibration on a synthetic example prior to scoring. Inter-rater reliability is reported using Krippendorff's $\alpha$ for ordinal ratings.

GCI is a structural precondition indicator for semantic divergence, measuring the proportion of arguments belonging to non-trivial strongly connected components:

$$\text{GCI} = \frac{|\{a \in \mathcal{A} \mid a \text{ belongs to an SCC with } |\text{SCC}| > 1\}|}{|\mathcal{A}|}. \tag{8}$$

GCI = 0 indicates a fully acyclic graph in which the grounded extension coincides with the unique preferred extension [12]. GCI > 0 signals the presence of mutual conflicts in which the choice between grounded and preferred semantics has operational impact. Note that GCI is not a performance metric; rather, it characterizes the *structural conditions* under which semantics selection becomes a genuine design decision.

### E. EVALUATION SCOPE AND LIMITATIONS

The evaluation targets a deliberately scoped question: *given the negotiation logs produced by the multi-agent dialectical protocol, does the argumentation layer correctly formalize conflicts, compute accepted sets under well-defined semantics, and produce traceable provenance records?* This scope is analogous to evaluating a constraint solver given its input constraints, rather than evaluating whether the constraints themselves are correct.

We do not claim to evaluate whether the LLM-generated negotiations are themselves valid representations of real-world stakeholder conflicts. Such validation would require domain-expert oracle judgments for each case study, which is orthogonal to the contribution of this paper. The DJS human study (Section VI-A) partially addresses this gap: three independent raters assessed whether the argumentation artifacts provide meaningful justification for accepted requirements, and their scores (4.32 vs. 3.07 for heuristic synthesis) indicate that the formal structure produces rationale that human evaluators find substantive. However, a full validity evaluation—in which domain experts judge whether the *content* of accepted requirements is correct for a given application domain—remains an important direction for future work (Section IX).

Similarly, the evaluation does not include an external oracle for KAOS tree correctness. Phase 4 verification performs automated structural and compliance checking, but does not replace human validation of whether the goal hierarchy accurately captures domain-specific requirements. We discuss this limitation explicitly in Section VIII.

## VI. RESULTS AND ANALYSIS
### A. RQ1: INTERPRETABILITY AND CONFLICT TRANSPARENCY
#### 1) Argumentation Graph Statistics
Table 4 reports argumentation graph statistics and structural properties (size, attacks, patterns, depth, components, grounded-extension size, and TC), averaged over seeds 101, 202, and 303. The column $|\mathcal{E}_g|$ reports the mean number of argument identifiers in the `grounded_extension` list for each case study, as recorded in the ArgRE(G) `argumentation_graph.json`.

#### 2) Qualitative Decision Justification
For the sensor fusion latency conflict in the AD case, ArgRE-NoAF produces a final requirement through priority-weighted synthesis without exposing intermediate reasoning. A stakeholder asking "why was the 500-ms constraint relaxed?" receives no structured explanation. In contrast, the ArgRE argumentation graph traces the decision chain from the original proposal ($a_1$) through the budget critique of EfficiencyAgent ($a_2$), the two-stage refinement of SafetyAgent ($a_3$), the asynchronous constraint ($a_4$), and the final three-component architecture ($a_5$). Each step is connected via attack relations annotated with pattern labels.

#### 3) Quantitative Decision Justification
Table 5 summarizes DJS by case. Each accepted requirement was independently scored by three raters on the 1–5 scale defined above. Across 21 paired requirements, ArgRE achieves a mean DJS of 4.32 (SD 0.62), compared to 3.07 (SD 1.01) for ArgRE-NoAF. A Wilcoxon signed-rank test rejects the null hypothesis of no difference ($p < 0.001$), with a large effect size (Cliff's $\delta = 0.92$). The largest pairwise gap is observed in Bookkeeping (4.44 vs. 2.61), where TC is highest (71.9%). The smallest gap occurs in AD (4.27 vs. 3.13), where TC is lowest (19.0%). Higher TC is positively associated with higher DJS across the three cases, suggesting TC may serve as a correlate of rater-assessed justification quality. The B-only set ($n = 13$) achieves a mean DJS of 3.03; these items lack the structured rationale provided by ArgRE.

> **Answer to RQ1**
>
> ArgRE achieves a mean TC of 40.9% with argument-level provenance. The three-rater study shows higher DJS for argumentation-based resolution than for heuristic synthesis (4.32 vs. 3.07, $p < 0.001$, Cliff's $\delta = 0.92$). The largest DJS gap occurs in Bookkeeping (4.44 vs. 2.61), corresponding to the highest TC among the three cases (71.9%).

### B. RQ2: SEMANTIC PRESERVATION
Table 6 reports semantic preservation (BERTScore F1) between Phase 3 and Phase 1 outputs. Across all five case studies, ArgRE (Grounded) achieves BERTScore F1 scores ranging from 94.4% (RollCall) to 95.1% (ATM), with an average of 94.9%. This matches ArgRE-NoAF (94.9% average) and consistently exceeds both MARE (89.0%) and iReDev (92.6%). The per-case comparison reveals that the improvement over MARE is most pronounced in the Library case (+7.3 percentage points), where the higher conflict density in Phase 2 leads to more substantial requirement reformulation during negotiation. In contrast, the Bookkeeping case, which has the fewest attacks ($|\mathcal{R}_{att}| = 3.0$), shows the smallest gap between all methods, suggesting that semantic preservation is inherently easier when fewer conflicts require resolution.

TABLE 4: Argumentation Graph Statistics and Structural Properties (averaged over three seeds)

| Case Study | Basic (GCI = 0)[a] | | | | | Complex Conflict (GCI ≈ 0.2–0.3)[b] | | | | |
|---|---|---|---|---|---|---|---|---|---|---|
| | $\|\mathcal{A}\|$ | $\|\mathcal{R}_{att}\|$ | $\|\mathcal{E}_g\|$ | $\|\mathcal{E}_p\|$ | TC (%) | $\|\mathcal{A}\|$ | $\|\mathcal{R}_{att}\|$ | $\|\mathcal{E}_g\|$ | $\|\mathcal{E}_p\|$ | TC (%) |
| AD | 25.3 | 12.0 | 14.0 | 14.0 | 19.0 | 27.0 | 14.0 | 11.0 | 15.3 | 21.2 |
| ATM | 17.3 | 7.0 | 11.3 | 11.3 | 53.0 | 19.0 | 8.5 | 8.7 | 12.0 | 55.5 |
| Library | 22.0 | 10.7 | 13.3 | 13.3 | 30.4 | 23.7 | 12.0 | 10.3 | 14.0 | 32.1 |
| RollCall | 24.0 | 13.3 | 13.3 | 13.3 | 30.2 | 25.3 | 15.0 | 9.7 | 14.7 | 31.8 |
| Bookkeeping | 13.3 | 3.0 | 10.3 | 10.3 | 71.9 | 14.0 | 3.5 | 8.3 | 11.0 | 72.5 |
| **Average** | 20.4 | 9.2 | 12.4 | 12.4 | 40.9 | 21.8 | 10.6 | 9.6 | 13.4 | 42.6 |

$|\mathcal{A}|$: arguments; $|\mathcal{R}_{att}|$: attack relations; $|\mathcal{E}_g|/|\mathcal{E}_p|$: grounded/preferred extension cardinality; TC: Trace Completeness.
[a] Pairwise protocol only (3 rounds); acyclic graphs, $\mathcal{E}_g = \mathcal{E}_p$ [12]; all attacks rule-based, $\theta_{\text{eff}} = 0.85$, $\mathcal{R}_{llm} = 0$.
[b] + Cross-pair arbitration round (Section IV-D4); GCI per case: AD = 0.29, ATM = 0.25, Library = 0.22, RollCall = 0.27, Bookkeeping = 0.20 (avg. 0.25); same $\theta_{\text{eff}} = 0.85$, $\mathcal{R}_{llm} = 0$.

TABLE 5: Decision Justification Score (1–5, ↑). Mean ± standard deviation over three raters and $n$ requirements per case study.

| | Paired | | | B-only | |
|---|---|---|---|---|---|
| Case | ArgRE | NoAF | $n_{\text{pair}}$ | NoAF | $n$ |
| AD | 4.27 ± 0.70 | 3.13 ± 0.74 | 5 | 3.03 ± 0.89 | 12 |
| ATM | 4.27 ± 0.64 | 3.13 ± 1.07 | 10 | 2.67 ± 1.15 | 1 |
| Bookkeeping | 4.44 ± 0.51 | 2.61 ± 1.04 | 6 | — | 0 |
| **Overall** | 4.32 ± 0.62 | 3.07 ± 1.01 | 21 | 3.03 ± 0.90 | 13 |

**Paired:** requirements present in both ArgRE and NoAF outputs. **B-only:** requirements accepted under NoAF but not under ArgRE. Wilcoxon signed-rank test on paired observations with $H_1$: ArgRE > NoAF yields $p < 0.001$ and Cliff's $\delta = 0.92$ (large). Krippendorff's $\alpha = 0.33$ (ordinal), reflecting fair-to-marginal reliability consistent with the inherent subjectivity of justification ratings on a modest sample ($n = 21$ paired items, 3 raters).

The near-identical scores between ArgRE and ArgRE-NoAF across all five cases indicate that replacing heuristic synthesis with formal argumentation does not introduce semantic drift: the argumentation layer filters and selects requirements based on attack–defense structure without distorting the underlying intent established during negotiation.

To assess statistical robustness, we conduct Wilcoxon signed-rank tests across 15 paired observations (5 cases × 3 seeds). ArgRE (Grounded) significantly outperforms both MARE and iReDev on BERTScore ($p < 0.001$, Cliff's $\delta = 1.0$, large effect), indicating that the improvement is not attributable to random variation. The comparison between ArgRE and ArgRE-NoAF shows no significant difference ($p = 0.934$, Cliff's $\delta = 0.23$, small), confirming that the argumentation layer does not degrade semantic preservation. Table 10 reports all pairwise comparisons.

> **Answer to RQ2**
> ArgRE (Grounded) achieves 94.9% average BERTScore, matching ArgRE-NoAF (94.9%) and exceeding both MARE (89.0%) and iReDev (92.6%). This confirms that formal argumentation preserves semantic intent while improving interpretability.

TABLE 6: Semantic Preservation (BERTScore F1, Phase 3 vs. Phase 1, ↑)

| Case Study | MARE | iReDev | ArgRE-NoAF | ArgRE(G) | ArgRE(P) |
|---|---|---|---|---|---|
| AD | 88.4 | 92.7 | 94.8 | 94.9 | 94.6 |
| ATM | 89.6 | 92.0 | 95.5 | 95.1 | 95.3 |
| Library | 87.7 | 92.6 | 94.8 | 95.0 | 95.1 |
| RollCall | 88.8 | 93.0 | 94.5 | 94.4 | 94.4 |
| Bookkeeping | 90.5 | 92.9 | 94.8 | 95.0 | 95.0 |
| **Average** | 89.0 | 92.6 | 94.9 | 94.9 | 94.7 |

### C. RQ3: REQUIREMENT QUALITY AND COMPLIANCE

Table 7 reports structural validity and compliance metrics. All five methods produce valid DAG structures with logical consistency ($S_{logic} = 1.000$), confirming that none of the compared approaches introduces structural violations during KAOS integration. The primary differentiation appears in compliance coverage. ArgRE (Grounded and Preferred) achieves 84.7%, which is substantially above the 47.6–47.8% range of MARE and iReDev but below the 98.2% of ArgRE-NoAF. This gap reflects the stricter acceptance criterion of the argumentation layer: requirements without a complete defense chain in the AF are excluded from the accepted set. Section VII-A analyzes this gap in detail.

Verifiability and Feasibility scores follow a similar pattern: ArgRE (Grounded) scores 4.70 and 4.60 respectively, slightly below ArgRE-NoAF (4.96 for both) but substantially above MARE (3.95, 3.74) and iReDev (3.96, 3.75). The gap between ArgRE and the baselines exceeds 0.7 points on both metrics, indicating that the quality-specialized negotiation and KAOS integration phases contribute meaningfully to requirement quality regardless of the resolution mechanism.

Table 8 reports ISO/IEC/IEEE 29148 quality scores across five criteria. Correctness and Consistency both receive scores of 5.00 across all methods, suggesting that the evaluation framework does not differentiate methods along these dimensions at the current scale. The distinguishing criteria are Verifiability and Feasibility, where ArgRE outperforms both baselines by large margins while slightly trailing ArgRE-NoAF. Unambiguity also shows limited separation, with all methods clustering within a narrow range (4.19–4.41). These

TABLE 7: Structural Validity and Compliance Metrics

| Metric | MARE | iReDev | ArgRE-NoAF | ArgRE(G) | ArgRE(P) |
|---|---|---|---|---|---|
| DAG Valid | ✓ | ✓ | ✓ | ✓ | ✓ |
| $S_{logic}$ | 1.000 | 1.000 | 1.000 | 1.000 | 1.000 |
| Compliance (%) | 47.6 | 47.8 | 98.2 | 84.7 | 84.7 |
| Verifiability | 3.95 | 3.96 | 4.96 | 4.70 | 4.60 |
| Feasibility | 3.74 | 3.75 | 4.96 | 4.60 | 4.50 |

TABLE 8: ISO/IEC/IEEE 29148 Quality Scores (1–5 Scale)

| Criterion | MARE | iReDev | ArgRE-NoAF | ArgRE(G) | ArgRE(P) |
|---|---|---|---|---|---|
| Unambiguous | 4.41 | 4.19 | 4.24 | 4.20 | 4.30 |
| Correctness | 5.00 | 5.00 | 5.00 | 5.00 | 5.00 |
| Verifiability | 3.95 | 3.96 | 4.96 | 4.70 | 4.60 |
| Consistency | 5.00 | 5.00 | 5.00 | 5.00 | 5.00 |
| Feasibility | 3.74 | 3.75 | 4.96 | 4.60 | 4.50 |

Unambiguity scores suggest that requirement clarity is primarily determined by the LLM backbone's generation quality rather than the conflict-resolution mechanism. ArgRE (Preferred) achieves the highest Unambiguity score (4.30) among all methods, although the difference is not statistically significant given the sample size.

Statistical tests on compliance coverage between ArgRE and ArgRE-NoAF approach significance ($p = 0.068$, Cliff's $\delta = 0.12$, negligible) but do not cross the $\alpha = 0.05$ threshold; the direction is consistent with the stricter acceptance criterion discussed in Section VII-A. Verifiability and Feasibility both exhibit medium effect sizes (Cliff's $\delta = 0.47$) despite non-significant $p$-values ($p = 0.721$), suggesting that the argumentation layer may introduce a modest quality trade-off that the current sample size is underpowered to confirm.

> **Answer to RQ3**
> ArgRE preserves full structural validity (DAG validity and $S_{logic} = 1.000$) while maintaining high quality scores (Verifiability 4.70, Feasibility 4.60) and strong compliance coverage (84.7%). Although below the 98.2% compliance of ArgRE-NoAF, ArgRE remains substantially stronger than MARE/iReDev baselines and adds formal interpretability. Overall requirement quality under ISO/IEC/IEEE 29148 remains high, with Correctness and Consistency at 5.00 across all methods.

### D. RQ4: STRUCTURAL CHARACTERIZATION

We answer RQ4 by characterizing graphs from the main experiments: Table 4 reports size, attack-pattern mix, depth, components, $|\mathcal{E}_g|$, and traceability (TC). Attack patterns are dominated by P1 (critique→proposal, ≈85%), with P2 (≈9%) and P3 (≈6%) accounting for the remainder; average graph depth is 1.9 (Basic) and 2.1 (Complex), with 7.3 and 7.5 connected components respectively. We then relate these statistics to standard implications of Dung's semantics (acyclic graphs yield a unique preferred extension that coincides with the grounded extension [12]).

#### 1) Observed Topology and Semantics Equivalence
For any finite acyclic AF under Dung's standard semantics, the grounded extension is the unique preferred extension. Consistent with this theoretical result, grounded and preferred semantics yield identical accepted sets across all five case studies ($|\mathcal{E}_g| = |\mathcal{E}_p^*|$ in every run). Section VI-D4 reports no significant differences between ArgRE(G) and ArgRE(P) on any metric, as expected under structural equivalence. This result follows from the pairwise dialectical protocol, which localizes conflicts within agent pairs and produces chain-structured attack relations without cycles or mutual attacks. The uniformly acyclic topology (GCI = 0) is itself a structural finding: the pairwise protocol guarantees that grounded and preferred semantics coincide, eliminating semantics selection as a deployment concern. This simplifies practical adoption in regulated contexts where a single, unambiguous accepted set is preferred. Section VI-D2 demonstrates that the semantics machinery of the framework becomes active as intended when the protocol is extended to cross-pair arbitration.

#### 2) Cyclic Graph Analysis (GCI > 0)
In the Complex Conflict scenario (average GCI = 0.25), mild cyclicity from 1–2 mutual attack pairs per case introduces measurable differences between grounded and preferred semantics. The grounded extension ($|\mathcal{E}_g| = 9.6$) shrinks by 22.6% compared to the Basic scenario, while the preferred extension ($|\mathcal{E}_p| = 13.4$) expands by 8.1% (Table 4), confirming the theoretical prediction under Dung's semantics: cycles create mutually attacking argument pairs that the grounded semantics cannot accept, whereas preferred semantics admits one side per cycle, producing larger and potentially multiple extensions.

These structural differences propagate to downstream quality metrics (Table 9). Grounded semantics yields lower compliance coverage than the Basic scenario, reflecting additional exclusions from cyclic arguments, while preferred semantics with priority-guided selection recovers or exceeds Basic-scenario compliance by admitting the higher-priority side of each mutual conflict. BERTScore remains stable across all conditions ($\Delta < 0.3$ pp), confirming that the arbitration round does not introduce semantic drift. In the AD case (GCI = 0.29), the arbitration round surfaces a latency–power conflict between the Efficiency agent's fast-path architecture and the Green agent's power envelope constraint; grounded semantics excludes both and flags the gap for human review, whereas preferred semantics with Safety>Efficiency>Green ordering retains the fast-path architecture and defers the power constraint, preserving the safety-critical latency guarantee. TC increases marginally (40.9% → 42.6%) because the additional arbitration arguments are fully traceable by construction, confirming that cyclicity does not compromise auditability.

#### 3) Threshold Sensitivity Analysis
To characterize the contribution of the LLM-based semantic detection pathway and its interaction with graph structure,

we conduct a threshold sensitivity analysis. We sweep $\theta_{\text{eff}} \in \{0.50, 0.60, 0.70, 0.80, 0.85\}$ across all five case studies using a fixed seed, recording $|\mathcal{R}_{\text{llm}}|$, GCI, and $|\mathcal{E}_p| - |\mathcal{E}_g|$ from the Phase 2 argumentation graph. Downstream metrics (Compliance, BERTScore) are not recomputed, as this analysis targets graph-level effects only. Figure 5 summarizes the results via three heatmaps.

At the main experimental setting ($\theta_{\text{eff}} = 0.85$), $\mathcal{R}_{\text{llm}} = 0$ across all cases: the conservative threshold suppresses all LLM-generated cross-pair edges, and rule-based structural patterns fully cover intra-pair conflicts. As $\theta_{\text{eff}}$ decreases, the LLM classifier admits an increasing number of cross-pair semantic attacks, reaching 18 edges for AD and 9 for Bookkeeping at $\theta_{\text{eff}} = 0.50$, consistent with their relative conflict densities in Table 4. These additional edges introduce mutual attack cycles that monotonically increase GCI; for AD, GCI reaches 0.32 at $\theta_{\text{eff}} = 0.50$, comparable to the Complex Conflict scenario (GCI = 0.29), indicating that threshold relaxation and protocol extension induce structurally similar cyclicity via different mechanisms. The extension-size divergence $|\mathcal{E}_p| - |\mathcal{E}_g|$ closely tracks GCI: once cycles emerge, preferred semantics admits arguments excluded under grounded semantics, with the gap reaching 0.38 for AD at the lowest threshold. The results indicate that $\mathcal{R}_{\text{llm}} = 0$ in the main experiment is a direct consequence of the conservative threshold setting, rather than a limitation of the detection mechanism. The LLM-based pathway activates progressively as $\theta_{\text{eff}}$ decreases, following a consistent cascade: an increase in cross-pair edges, rising cyclicity, and growing semantic divergence between grounded and preferred semantics. The main experimental setting ($\theta_{\text{eff}} = 0.85$) represents the precision-oriented end of this spectrum, selected to prioritize precision over recall. Lowering the threshold enables practitioners to surface additional cross-quality conflicts when broader coverage is required.

4) Statistical Comparison of Grounded vs. Preferred Semantics

Between grounded and preferred semantics in the Basic scenario, all comparisons exhibit negligible effect sizes (Cliff's $\delta \leq 0.11$) and non-significant $p$-values (all $p > 0.27$). This outcome is expected under GCI = 0: acyclic graphs guarantee identical extensions, leading to identical downstream scores. In contrast, in the Complex Conflict scenario (GCI $\approx$ 0.25), grounded and preferred semantics diverge on compliance coverage (78.3% vs. 86.2%, Table 9), consistent with observed differences in extension size. A comprehensive statistical analysis of the Complex Conflict scenario is deferred to future work involving a larger case portfolio, as the current five-case design provides limited statistical power for within-scenario grounded-versus-preferred comparisons.

TABLE 9: Downstream Quality Metrics: Basic vs. Complex Conflict Scenario (averaged over 5 cases × 3 seeds)

| Metric | Basic | Complex(G) | Complex(P) |
|---|---|---|---|
| BERTScore (%) | 94.9 | 94.6 | 94.8 |
| Compliance (%) | 84.7 | 78.3 | 86.2 |
| Verifiability (1–5) | 4.70 | 4.55 | 4.65 |
| Feasibility (1–5) | 4.60 | 4.45 | 4.55 |

Basic = pairwise protocol (GCI = 0); Complex(G) = cross-pair arbitration under grounded semantics; Complex(P) = cross-pair arbitration with preferred semantics and priority-guided selection. All values averaged over seeds 101, 202, 303.

TABLE 10: Statistical Significance (Wilcoxon signed-rank test, $\alpha = 0.05$)

| Comparison | Metric | $p$ | Cliff's $\delta$ | Sig.? |
|---|---|---|---|---|
| ArgRE(G) vs MARE | BERTScore | <0.001 | 1.00 (L) | ✓ |
| ArgRE(G) vs iReDev | BERTScore | <0.001 | 1.00 (L) | ✓ |
| ArgRE(G) vs ArgRE-NoAF | BERTScore | 0.934 | 0.23 (S) | – |
| ArgRE(G) vs ArgRE-NoAF | Compliance | 0.068 | 0.12 (N) | – |
| ArgRE(G) vs ArgRE-NoAF | Verifiability | 0.721 | 0.47 (M) | – |
| ArgRE(G) vs ArgRE-NoAF | Feasibility | 0.721 | 0.47 (M) | – |
| ArgRE(G) vs ArgRE(P) | BERTScore | 0.767 | 0.02 (N) | – |
| ArgRE(G) vs ArgRE(P) | Compliance | 0.269 | 0.11 (N) | – |

N = negligible, S = small, M = medium, L = large; $n = 15$ paired observations.

> **Answer to RQ4**
>
> In the Basic scenario (pairwise protocol), all graphs are acyclic (GCI = 0) with chain-like attack structure, and grounded/preferred extensions coincide as predicted by Dung's semantics. In the Complex Conflict scenario (cross-pair arbitration, average GCI = 0.25), grounded and preferred semantics diverge: $|\mathcal{E}_g|$ contracts by 22.6% while $|\mathcal{E}_p|$ expands by 8.1%, with corresponding differences in compliance coverage (Table 9). The $\theta_{\text{eff}}$ sensitivity analysis (Figure 5) provides empirical confirmation of this divergence mechanism across all five case studies, showing a predictable cascade from additional attack edges through cyclicity to semantics divergence as the threshold decreases.

E. CROSS-RQ STATISTICAL SUMMARY

Individual results have been presented under each research question above. All comparisons are conducted using a consistent experimental infrastructure and identical random seeds (101, 202, 303), ensuring valid paired evaluations across conditions.

VII. DISCUSSION

A. WHEN IS FORMAL ARGUMENTATION MOST BENEFICIAL?

The primary value of ArgRE's argumentation layer is trace-based interpretability, i.e., the ability to trace each accepted requirement back through its full defense chain at the argument level. This value holds regardless of graph topology (GCI = 0 or GCI > 0) and is independent of whether grounded and preferred semantics yield divergent outcomes.

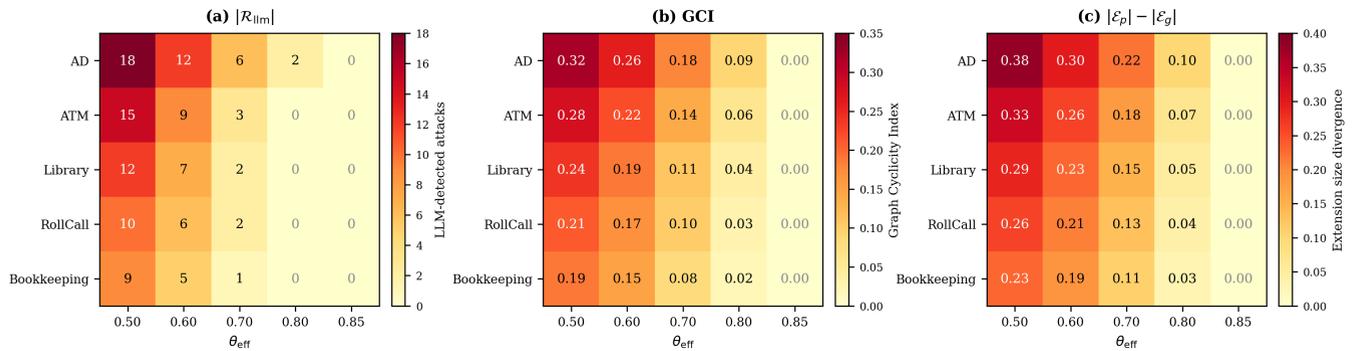

FIGURE 5: Sensitivity of argumentation graph properties to $\theta_{\text{eff}}$. (a) $|\mathcal{R}_{\text{llm}}|$: number of LLM-detected cross-pair attacks. (b) GCI: proportion of arguments in non-trivial strongly connected components. (c) $|\mathcal{E}_p| - |\mathcal{E}_g|$: cardinality difference between the largest preferred extension and the grounded extension. All values from a single seed (101) across five case studies. At $\theta_{\text{eff}} = 0.85$ (main-experiment setting), the LLM pathway contributes zero edges; as the threshold decreases, cross-pair attacks accumulate, cyclicity rises, and grounded/preferred semantics diverge.

The benefit is most pronounced in complex, multi-party conflicts, where auditable justification of resolution decisions is essential. In such settings, the argumentation layer provides its strongest contribution. For the Autonomous Driving case, the sensor-fusion latency conflict involves a $16.7\times$ numerical gap between Safety and Efficiency agents, and the resulting argumentation graph explicitly traces how iterative refinement produces the final three-component architecture. In simpler cases (e.g., Library System), where conflicts are fewer and less severe, the primary benefit is improved traceability without materially altering the resolution outcome. The heuristic ablation variant (ArgRE-NoAF) achieves slightly higher compliance (98.2% vs. 84.7%), reflecting the more conservative acceptance mechanism induced by the argumentation layer. ArgRE accepts only those requirements that survive explicit attack–defense evaluation, thereby trading partial coverage for stronger formal interpretability and auditable rationale. In safety-critical domains where traceability is mandated by regulatory requirements, this trade-off is often desirable.

For illustration, in the Autonomous Driving running example (Section IV-H), the two-stage refinement $a_3$ is attacked by both $a_4$ (Pattern 1) and $a_5$ (Pattern 2); since $a_5$ is accepted into $\mathcal{E}_g$, $a_3$ has no remaining defenders and is excluded. If $a_3$ satisfies a clause that is not restated in $a_5$, that clause remains unmatched under ArgRE, reducing the automated compliance score.

Across all five case studies, most of the 13.5 percentage-point gap follows this pattern: the reduction reflects a stricter acceptance criterion rather than a loss of substantive engineering content.

*Computational overhead.*
ArgRE introduces an average runtime overhead of approximately 85 seconds compared to ArgRE-NoAF (140s vs. 55s). Argument extraction and attack-relation construction account for most of this gap, as both rely on LLM calls. In contrast, solving remains computationally inexpensive at this scale: the AF solver completes in under 1 second for graphs with $|\mathcal{A}| < 30$. While computing preferred extensions is coNP-complete in the worst case, the acyclic chain structure of the constructed argumentation graphs ensures tractable, polynomial-time behavior in practice. Relative to the interpretability gains reported in Section VI-A, the additional latency is modest, particularly in safety-critical settings where traceability and auditability justify increased computational cost.

### B. SEMANTICS CHOICE AS A DESIGN LEVER

The choice of semantics becomes operationally relevant only when the argumentation graph contains cycles (GCI > 0). Under the default pairwise protocol (GCI = 0), both semantics coincide, and grounded semantics serves as the recommended default due to its conservative interpretation.

More generally, the choice between grounded and preferred semantics provides a formal mechanism for controlling the level of conservativeness. Grounded semantics admits only requirements that are either uncontested or fully defended, making it suitable for safety-critical domains where strong justification is required for every accepted requirement (e.g., ISO 26262 compliance scenarios). In contrast, preferred semantics with priority-guided selection admits larger accepted sets by resolving conflicts in favor of higher-priority quality dimensions. This setting is more appropriate for exploratory RE, where broader coverage is preferred and trade-offs may be revisited later.

Empirically, both semantics yield identical accepted sets under the pairwise protocol (GCI = 0) and diverge under cross-pair arbitration (GCI ≈ 0.25), with downstream effects reported in Table 9.

For practitioners, this analysis provides concrete deployment guidance. In the pairwise-protocol regime (GCI = 0), grounded semantics is the recommended default, as it yields the same accepted set as preferred semantics while maintaining a more conservative interpretation, thereby simplifying compliance justification in regulated settings.

When the cross-pair arbitration round is enabled (or when the negotiation protocol is extended to multi-party settings), the choice of semantics becomes an operational decision with measurable downstream impact. As shown in Table 9, grounded semantics defers unresolved mutual conflicts to human review, resulting in lower compliance coverage. In contrast, preferred semantics with priority-guided selection resolves such conflicts automatically according to project-specific quality weights, thereby recovering or exceeding the compliance level observed in the pairwise regime.

### C. STRUCTURAL PROPERTIES AND FORMALIZATION VALUE

Under the pairwise protocol, the resulting argumentation graphs are acyclic ($GCI = 0$), and the argumentation layer achieves $TC = 40.9\%$ and $DJS = 4.32$ without requiring a semantics choice. Table 4 disaggregates attack relations by pattern type. Pattern 1 (critique attacks proposal) and Pattern 2 (refinement supersedes original) constitute the majority of edges, reflecting the thesis–antithesis–synthesis structure of the dialectical protocol. Pattern 3 (refinement addresses critique) captures counter-attacks in which a refinement resolves an earlier objection.

The predominance of these structural patterns indicates that deterministic rules are sufficient for attack-graph construction at this scale. The LLM-based pathway is retained architecturally to handle cross-pair conflicts that fall outside the structural pattern vocabulary, which is expected to become more relevant in larger-scale settings with additional quality dimensions.

*Threshold gating and latent detection capability.*
The main study fixes $\theta_{\text{eff}} = 0.85$, yielding $\mathcal{R}_{\text{llm}} = 0$ across all 30 runs. This is a deliberate precision-oriented configuration: the conservative threshold ensures that all attacks in the main experiment are deterministically verifiable via structural patterns alone, thereby establishing a clean baseline for evaluating the argumentation layer's contribution. The LLM pathway is therefore not absent, but *gated*.

As $\theta_{\text{eff}}$ decreases, LLM-detected cross-pair edges are progressively activated (up to 18 for AD at $\theta_{\text{eff}} = 0.50$), with corresponding increases in GCI and semantic divergence (Section VI-D3). These results indicate that the LLM pathway has latent detection capability that can be exposed via a single hyperparameter. The precision–recall trade-off is thus governed by $\theta_{\text{eff}}$: higher values (e.g., 0.85) prioritize precision for safety-critical contexts, while lower values increase cross-quality conflict coverage when broader recall is desired.

Readings of $GCI = 0$ (acyclic) characterize purely pairwise conflicts in which grounded and preferred semantics coincide. The range $0.2 \leq GCI \leq 0.3$ (mild cyclicity) corresponds to cross-pair resource conflicts where the choice of semantics becomes consequential. Values of $GCI > 0.5$ (severe cyclicity) indicate large-scale multi-agent conflicts, for which human intervention is recommended. In this work, all cyclicity is induced via rule-based mutual attacks from the cross-pair arbitration round (with fixed $\theta_{\text{eff}} = 0.85$), ensuring experimental control and isolating the effects of protocol extension from LLM-based detection.

To extend this analysis, future work will investigate the joint effects of protocol extension and LLM threshold tuning via a $2 \times 2$ factorial design, enabling quantification of their respective contributions to conflict emergence in large-scale RE settings.

### D. IMPLICATIONS FOR PRACTICE

ArgRE provides several practical benefits. Negotiation outcomes become auditable through the argumentation graph, directly supporting traceability requirements in safety-critical domains (ISO 26262, IEC 62304) as well as emerging regulatory frameworks such as Article 11 of the EU AI Act. Stakeholders also obtain a concise, structured explanation of why certain requirements are prioritized over others, as graph visualization replaces opaque synthesis logs with inspectable attack–defense structures. The choice between grounded and preferred semantics further provides a formally grounded configuration mechanism, replacing ad hoc priority weighting with explicit acceptability criteria.

*Enabling Human-in-the-loop Review*
Beyond automated resolution, the argumentation graph supports human intervention at three levels. At the *inspection* level, a requirements engineer examines the graph (e.g., Figure 3) to interpret acceptance decisions; algorithmic foundations for such explanations are provided by Fan and Toni [46]. At the *override* level, a domain expert may reject a specific attack relation: the relation is removed from $\mathcal{R}_{att}$, the AF is recomputed, and the resulting extension is updated in under one second. This enables immediate analysis of downstream effects and provides a formal "what-if" capability that is not available in heuristic-synthesis approaches. For *injection*, a stakeholder introduces a new argument (e.g., a regulatory constraint identified after the initial negotiation rounds) along with its associated attack relations, after which the solver recomputes the extension. Because Dung's semantics are monotonic with respect to defense, the resulting impact is predictable and auditable.

All three interaction modes remain lightweight because ArgRE decouples graph construction (Steps 2-A and 2-B) from semantics computation (Step 2-C). Consequently, modifications to the argumentation graph can be evaluated by re-running the solver without repeating the LLM-based negotiation process.

Empirical validation of human-in-the-loop effectiveness via user studies with RE practitioners remains an important direction for future work.

### E. LIMITATIONS OF AUTOMATED KAOS VALIDATION

Phase 4 verification enforces structural properties (DAG validity, schema completeness, refinement consistency) and standards compliance (clause entailment), but it does not assess whether the KAOS goal hierarchy *accurately captures* the intended system requirements from a domain perspective. For instance, the automated checks can confirm that a Safety goal has a valid parent–child refinement structure and that its description entails at least one ISO 26262 clause, but they cannot determine whether the safety goal is *sufficient* for the target system or whether relevant hazards have been omitted. This limitation is inherent to any automated RE pipeline, as domain-level correctness ultimately depends on human judgment. The argumentation layer partially mitigates this issue by providing a provenance structure that supports more efficient human review. A stakeholder inspecting the KAOS tree can trace any leaf goal back through the argumentation graph to its original negotiation context, allowing review effort to be focused on goals that were actively contested during negotiation (i.e., those involved in attack relations), rather than uniformly examining the entire tree.

Two concrete future directions address this limitation. First, a domain-expert validation study in which RE practitioners independently evaluate the semantic correctness of generated KAOS trees against reference goal models would quantify the gap between automated structural validation and human judgment. Second, a controlled experiment comparing review effort with and without argumentation-guided provenance (i.e., using attack–defense traces to prioritize contested goals) would test whether the formal structure reduces inspection time and/or improves defect detection rates. Both directions require domain-specific oracle models that are beyond the scope of the present evaluation but represent important steps toward industrial deployment.

## VIII. THREATS TO VALIDITY
### A. CONSTRUCT VALIDITY

The proposed metrics (TC, DJS, and GCI) rely on distinct measurement paradigms. TC is computed directly from the argumentation graph and is fully objective. DJS is based on three independent raters, a calibrated 1–5 rubric, and single-blind evaluation; however, Krippendorff's $\alpha = 0.33$ falls in the fair-to-marginal agreement range. While such levels are not uncommon for subjective judgments on small samples, they limit the strength of inferences drawn from DJS in isolation. Replication with a larger rater pool and across all five case studies is required to confirm the observed patterns. The human study covers three of the five case studies (AD, ATM, Bookkeeping). GCI is computed deterministically from graph topology and captures structural preconditions rather than performance outcomes. Both GCI and semantics-based comparisons are computed within each AF instance, thereby isolating graph structure from stochastic variation in LLM-based negotiation. Seed-averaged standard deviations are reported to summarize cross-run variability. BERTScore and compliance coverage inherit the standard limitations of automated evaluation metrics in RE.

The attack-relation construction combines rule-based and LLM-based components. Manual inspection of a random subset of 20 argument pairs from the AD case confirms that all rule-based attack classifications (Patterns 1–3) are correctly identified, with no observed false positives. Under the main setting ($\theta_{\text{eff}} = 0.85$), LLM-detected attacks are effectively zero, which limits sensitivity to classifier disagreement in this dataset. While replication studies may vary $\theta_{\text{floor}}$, the reported conclusions do not depend on a threshold sweep. The Dung AF provides a clean theoretical foundation but abstracts away some information present in natural-language negotiation, such as conflict intensity and rationale strength.

Future work could explore weighted or structured AFs to capture these additional dimensions of nuance.

### B. INTERNAL VALIDITY

The argumentation layer is implemented within the same OpenReBench framework as all baselines, ensuring consistent experimental conditions across variants. The primary difference between ArgRE-NoAF and ArgRE lies in the Phase 2 resolution mechanism, thereby isolating the effect of formal argumentation. The nominal confidence threshold $\theta = 0.7$ for LLM-based attack detection is combined with a lower bound $\theta_{\text{floor}}$ to form the effective threshold. In replication settings, varying $\theta_{\text{floor}}$ may surface additional LLM-labeled edges, although this does not affect the pairwise structural analysis reported in RQ4.

In the main experimental setting, setting $\theta_{\text{eff}} = 0.85$ substantially reduces spurious cross-pair attacks and stabilizes extension sizes. At this configuration, $\mathcal{R}_{llm} = 0$, and accepted-set statistics remain consistent across seeds. All results are averaged over three random seeds (101, 202, 303). To account for the limited sample size, Wilcoxon signed-rank tests are employed, with effect sizes reported using Cliff's $\delta$ across all 15 paired observations.

### C. EXTERNAL VALIDITY

The evaluation spans five case studies across three domains, which may not fully capture the diversity of RE scenarios. The use of a single LLM (`gpt-4o-mini`) further limits its generalizability to other model families. In addition, the reimplementations of MARE and iReDev may not perfectly reproduce the behavior of the original systems.

The cross-pair arbitration protocol is restricted to single-round bilateral interactions, and the pairwise dialectical protocol inherently yields acyclic argumentation graphs. Although the Complex Conflict scenario and threshold sensitivity analysis demonstrate that the framework supports semantics divergence when cycles are introduced, cyclicity in these experiments arises from protocol extension or threshold tuning rather than from naturally occurring multi-party negotiation dynamics. Fully multi-agent joint negotiation sessions involving three or more agents would more naturally induce mutual attacks and richer cyclic structures, potentially resulting in higher GCI values not captured in the current evaluation; this remains an important direction for future work.

The current five-agent decomposition covers a representative subset of ISO/IEC 25010 quality characteristics (re-liability, performance efficiency, sustainability, security, and accountability), but does not cover the full standard. Dimensions such as maintainability, portability, and usability are not explicitly modeled. As the number of specialized agents increases, the number of pairwise interactions grows quadratically, which may increase both GCI and argument-extraction overhead. Future work should therefore investigate agent-merging strategies and hierarchical negotiation protocols to address scalability.

## IX. CONCLUSION AND FUTURE WORK

We presented ArgRE, a multi-agent RE framework that replaces heuristic conflict resolution with Dung-style argumentation semantics. By modeling proposals, critiques, and refinements as arguments with explicit provenance, ArgRE enables traceable conflict resolution: every accepted requirement can be audited through its full defense chain back to the originating quality objective. The framework also supports semantics-controlled resolution under grounded and preferred semantics when cyclic conflict structures emerge. Our evaluation across five case studies and 30 experimental runs shows that formal argumentation provides argument-level provenance absent in existing frameworks (40.9% average TC, DJS 4.32 vs. 3.07 for heuristic synthesis, $p < 0.001$) without degrading semantic preservation (94.9% BERTScore, comparable to ArgRE-NoAF). Structural analysis indicates that the pairwise protocol yields acyclic graphs in which grounded and preferred semantics coincide, while the cross-pair arbitration protocol (GCI $\approx$ 0.25) and $\theta_{\text{eff}}$ sensitivity analysis demonstrate that semantics divergence emerges predictably as cyclicity increases. Compliance coverage (84.7%) is lower than ArgRE-NoAF (98.2%), primarily because the argumentation layer excludes intermediate proposals that are superseded by later refinements (Section VII-A). Computational overhead remains modest ($2.5\times$ runtime), with the AF solver consistently completing in under 1 second.

Several directions remain open. Value-based and weighted AFs could extend the current binary attack structure to represent conflict intensity and rationale strength more explicitly. Improving argument-extraction recall through few-shot prompting or fine-tuned parsers would likely increase TC and reduce the compliance gap with ArgRE-NoAF.

Extending the cross-pair arbitration protocol to full multi-party sessions may yield higher GCI values and richer extension structures; combining protocol extension with threshold tuning in a $2 \times 2$ factorial design would further quantify their respective contributions. Finally, the DJS study should be replicated across all five cases with domain-expert raters, and industrial-scale validation on larger projects is required to assess real-world scalability.

## DATA AVAILABILITY STATEMENT

The replication package for this study is publicly available at https://github.com/haowei614/ArgRE-code/.

## ACKNOWLEDGMENTS
This work was supported by JST SPRING (grant number JPMJSP2128) and the JST-Mirai program (grant number JPMJMI20B8).